\documentclass[twocolumn,notitlepage,prx,superscriptaddress,longbibliography]{revtex4-2}
\usepackage{listings}
\usepackage{tikz}
\usepackage{times}
\usepackage{graphicx}
\usepackage{float}
\usepackage{latexsym,amsmath,amssymb,bm,euscript}
\usepackage{color}
\usepackage{subfigure}
\usepackage{epstopdf}
\usepackage[colorlinks=true,linkcolor=blue,citecolor=blue]{hyperref}
\usepackage{soul}
\usepackage[normalem]{ulem}
\usepackage{mathrsfs}
\usepackage{amssymb}
\usepackage{algpseudocode}
\usepackage{algorithm}
\usepackage{amsmath}
\usepackage{braket}
\usepackage{booktabs}
\usepackage{chngcntr}
\usepackage{lettrine}
\usepackage{lipsum}
\usepackage{bbding}
\usepackage{xspace}
\usepackage{textcomp}
\usepackage{textcase}
\usepackage{setspace}
\usepackage[T1]{fontenc}
\usepackage{float}
\usepackage{graphicx}
\usepackage{multirow}
\usepackage{mathtools}
\usepackage{threeparttable}
\usepackage[vlined, ruled, algo2e, linesnumbered]{algorithm2e}
\usepackage{changes}
\usepackage{bibunits}
\usepackage{natbib}
\usetikzlibrary{shapes}
\usepackage{url}

\def\CSO{CuSO$_4\cdot$5H$_2$O\xspace}
\def\CPN{CuPzN\xspace}
\def\CCC{(C$_5$H$_{12}$N)$_2$CuCl$_4$\xspace}
\def\CMN{Ce$_2$Mg$_3$(NO$_3$)$_{12}\cdot$24H$_2$O\xspace}

\def\para{\ensuremath{/\kern -0.8em /}\xspace}
\def\beqn{\begin{eqnarray}}
	\def\eeqn{\end{eqnarray}}\def\beq{\begin{equation}}
	\def\eeq{\end{equation}}

\newcommand{\Beq}{\begin{eqnarray*} }
	\newcommand{\Eeq}{\end{eqnarray*} }
\newcommand{\Bmat}{\left(\begin{matrix}}
	\newcommand{\Emat}{\end{matrix}\right)}

\makeatletter
\renewcommand{\fnum@figure}{\textbf{Figure.~\thefigure}}
\renewcommand{\fnum@table}{\textbf{Table \thetable}}
\makeatother

\graphicspath{{./Figs/}}

\begin{document}
\title{Universal Magnetocaloric Effect near Quantum Critical Point of\\ Magnon Bose-Einstein Condensation}

\author{Junsen Xiang}
\thanks{These authors contributed equally to this work.}
\affiliation{Beijing National Laboratory for Condensed Matter Physics, 
	Institute of Physics, Chinese Academy of Sciences, Beijing 100190, China}

\author{$\hspace{-1.4mm}^{,\, \dag}$~Enze Lv}
\thanks{These authors contributed equally to this work.}
\affiliation{Institute of Theoretical Physics, Chinese Academy of Sciences, 
	Beijing 100190, China}
\affiliation{School of Physical Sciences, University of Chinese Academy of 
Sciences, Beijing 100049, China}

\author{Qinxin Shen}
\thanks{These authors contributed equally to this work.}
\affiliation{Beijing National Laboratory for Condensed Matter Physics, 
	Institute of Physics, Chinese Academy of Sciences, Beijing 100190, China}
\affiliation{School of Physical Sciences, University of Chinese Academy of 
Sciences, Beijing 100049, China}

\author{Cheng Su}
\affiliation{School of Physics, Beihang University, Beijing 100191, China}

\author{Xuetong He}
\affiliation{Key Laboratory of Rare Earths, Ganjiang Innovation Academy, 		
	Chinese Academy of Sciences, Ganzhou, 341119, China}
\affiliation{School of Rare Earths, University of Science and Technology of China,
	Hefei, 230026, China}

\author{Yinghao Zhu}
\affiliation{Institute of Applied Physics and Materials Engineering, 
	University of Macau, Avenida da Universidade, Taipa, Macao SAR 999078, China}

\author{Yuan Gao}
\affiliation{School of Physics, Beihang University, Beijing 100191, China}
\affiliation{Institute of Theoretical Physics, Chinese Academy of Sciences, 
	Beijing 100190, China}

\author{Xin-Yang Liu}
\affiliation{School of Physics, Beihang University, Beijing 100191, China}
\affiliation{Beijing National Laboratory for Condensed Matter Physics, 
	Institute of Physics, Chinese Academy of Sciences, Beijing 100190, China}

\author{Dai-Wei Qu}
\affiliation{Kavli Institute for Theoretical Sciences, and School of Physical Sciences, 
	University of Chinese Academy of Sciences, Beijing 100049, China}
\affiliation{Institute of Theoretical Physics, Chinese Academy of Sciences, 
	Beijing 100190, China}

\author{Xinlei Wang}
\affiliation{School of Physics, Beihang University, Beijing 100191, China}

\author{Xi Chen}
\affiliation{Lanzhou Center for Theoretical Physics, Key Laboratory of Theoretical Physics of 
Gansu Province, Key Laboratory of Quantum Theory and Applications of MoE, Lanzhou University, 
Lanzhou 730000, China}

\author{Qian Zhao}
\affiliation{Institute of Applied Physics and Materials Engineering, 
University of Macau, Avenida da Universidade, Taipa, Macao SAR 999078, China}

\author{Haifeng Li}
\affiliation{Institute of Applied Physics and Materials Engineering, 
University of Macau, Avenida da Universidade, Taipa, Macao SAR 999078, China}

\author{Shuo Li}
\affiliation{Beijing National Laboratory for Condensed Matter Physics, 
	Institute of Physics, Chinese Academy of Sciences, Beijing 100190, China}

\author{Jie Yang}
\affiliation{Beijing National Laboratory for Condensed Matter Physics, 
Institute of Physics, Chinese Academy of Sciences, Beijing 100190, China}

\author{Jun Luo}
\affiliation{Beijing National Laboratory for Condensed Matter Physics, 
Institute of Physics, Chinese Academy of Sciences, Beijing 100190, China}

\author{Peijie Sun}
\affiliation{Beijing National Laboratory for Condensed Matter Physics, 
Institute of Physics, Chinese Academy of Sciences, Beijing 100190, China}

\author{Wentao Jin}
\email{Corresponding authors: xiangjs@iphy.ac.cn; wtjin@buaa.edu.cn; qiyang@fudan.edu.cn; rzhou@iphy.ac.cn; w.li@itp.ac.cn; sugang@itp.ac.cn}
\affiliation{School of Physics, Beihang University, Beijing 100191, China}

\author{{Yang Qi}}
\email{Corresponding authors: xiangjs@iphy.ac.cn; wtjin@buaa.edu.cn; qiyang@fudan.edu.cn; rzhou@iphy.ac.cn; w.li@itp.ac.cn; sugang@itp.ac.cn}
\affiliation{State Key Laboratory of Surface Physics and Department of Physics, 
Fudan University, Shanghai 200433, China}

\author{Rui Zhou}
\email{Corresponding authors: xiangjs@iphy.ac.cn; wtjin@buaa.edu.cn; qiyang@fudan.edu.cn; rzhou@iphy.ac.cn; w.li@itp.ac.cn; sugang@itp.ac.cn}
\affiliation{Beijing National Laboratory for Condensed Matter Physics, Institute of Physics, 
Chinese Academy of Sciences, Beijing 100190, China}
\affiliation{School of Physical Sciences, University of Chinese Academy of 
Sciences, Beijing 100049, China}

\author{Wei Li}
\email{Corresponding authors: xiangjs@iphy.ac.cn; wtjin@buaa.edu.cn; qiyang@fudan.edu.cn; rzhou@iphy.ac.cn; w.li@itp.ac.cn; sugang@itp.ac.cn}
\affiliation{Institute of Theoretical Physics, Chinese Academy of Sciences, Beijing 100190, China}
\affiliation{School of Physical Sciences, University of Chinese Academy of Sciences, Beijing 100049, China}

\author{Gang Su}
\email{Corresponding authors: xiangjs@iphy.ac.cn; wtjin@buaa.edu.cn; qiyang@fudan.edu.cn; rzhou@iphy.ac.cn; w.li@itp.ac.cn; sugang@itp.ac.cn}
\affiliation{Institute of Theoretical Physics, Chinese Academy of Sciences, Beijing 100190, China}
\affiliation{Kavli Institute for Theoretical Sciences, and School of Physical Sciences, 
	University of Chinese Academy of Sciences, Beijing 100049, China}

\begin{abstract} \bfseries \boldmath
Bose-Einstein condensation (BEC), a macroscopic quantum phenomenon arising from phase coherence and bosonic statistics, has been realized in quantum magnets. Here, we report the observation of a universal magnetocaloric effect (MCE) near a BEC quantum critical point (QCP) in copper sulfate crystal (\CSO). By conducting magnetocaloric and nuclear magnetic resonance measurements, we uncover a field-driven BEC QCP, evidenced by the universal scaling law $T_c \propto (B_c - B)^{2/3}$ and the perfect data collapse of the magnetic Grüneisen ratio. Thermal excitation triggers a dimensional crossover to a 1D quantum-critical regime, where the MCE scaling strictly matches the universality class of 1D Fermi gases. Notably, the quantum-critical MCE enables cooling down to 12.8~mK without helium-3, with very fast thermal relaxation rate that is critical for high cooling power. This work demonstrates the universal MCE in magnon BEC systems, using a common copper sulfate compound as a paradigmatic example, and paves the way for next-generation sub-Kelvin cooling.
\end{abstract}

% ====== Fig. 1 ====== %
\begin{figure*}[htp]
	\includegraphics[width=1\linewidth]{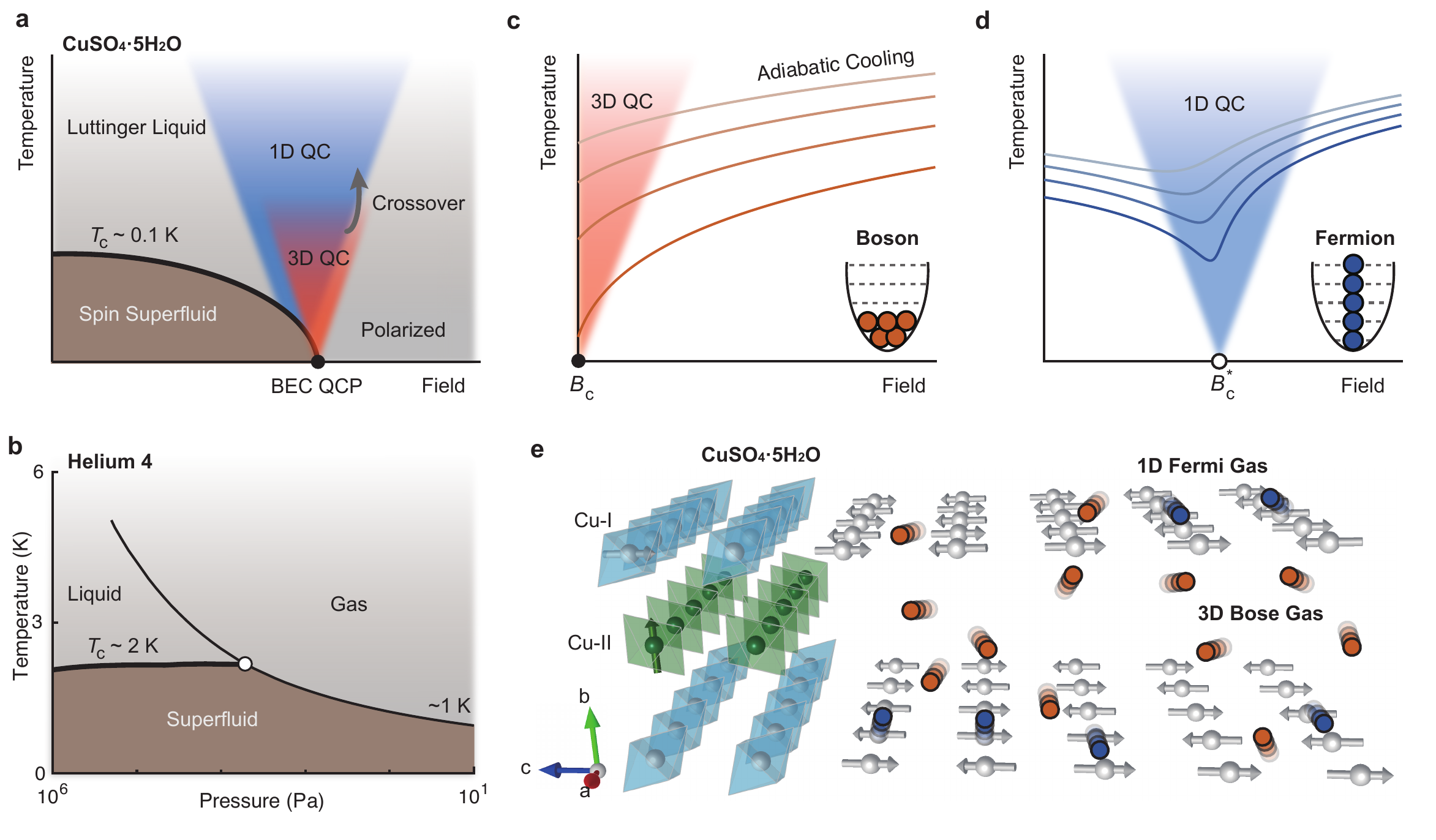}
	\caption{\textbf{Quantum critical regimes with universal MCE.} 
	\textbf{a} The schematic phase diagram of CSO under magnetic field. The black solid line 
	represents the superfluid transition $T_c \lesssim 0.1$~K, terminating at a 3D BEC QCP. The 
	orange fan shows a 3D QC regime, while the blue fan represents a 1D QC, effectively described by 3D 
	Bose and 1D Fermi gas theories, respectively.
	\textbf{b} The pressure-temperature phase diagram of helium-4. The superfluid transition (bold line) 
	occurs above 1~K. The thin line represents a first-order phase transition, 
	while the open circle marks a tricritical point. Panels 
	\textbf{c} and \textbf{d} display the isentropic lines for the ideal 3D Bose gas and 1D Fermi gas, respectively. 
	$B_c$ and $B_c^*$ represent the critical fields (or chemical potentials) of the two cases.
	\textbf{e} The magnetic structure of CSO, where Cu-I ions (gray balls) form spin chains along 
	the $a$-axis, while Cu-II ions (green balls) carry nearly free spins. The blue circles denote the fermions 
	(spinons), which propagate along the 1D chain, while the orange circles represent the bosons (magnons)
	moving in the 3D lattice. 
	} 
	\label{Fig1}
\end{figure*}
	
\date{\today}
\maketitle
\noindent
{\textbf{Introduction}}\\
Bose-Einstein condensate (BEC) is a quantum state of matter in which bosonic particles (e.g., 
atoms, quasiparticles like magnons and excitons) occupy the same lowest-energy quantum state
\cite{Anderson1995,Davis1995,Ensher1996,Dalfovo1999,Kasprzak2006,Bloch2008}, and exhibit 
macroscopic quantum coherence. BECs exist in fundamental quantum phenomena such as 
superfluidity, and also offer potential applications in quantum simulation~\cite{Gross2017Simulations} 
and information technologies~\cite{Treutlein2018RMP}. Magnon BECs, in particular, represent a 
unique class of quantum states in magnetic systems that exhibit rich spin-dependent phenomena 
and field-tunable quantum effects~\cite{Giamarchi2008NP,Batista2014RMP}. The BEC transition
and associated quantum critical point (QCP) emerge in various quantum magnetic systems, including 
the dimer systems~\cite{TlCuCl3_2003N,Jaime2004HanPurple,Sebastian2006Purple}, coupled 
spin chains~\cite{Zapf2006DTN,Yin2008BEC}, and 2D spin lattices~\cite{Xiang2024Nature,
Matsumoto2024BEC,Wu2022PNAS,Gao2022QMats}, etc. While most such systems require critical 
fields exceeding 10 T, the exploration of highly tunable BEC QCP at moderate magnetic fields 
represents a compelling topic in quantum magnetism.

Copper sulfate pentahydrate (\CSO), abbreviated CSO, is a multifunctional crystal with broad applications 
in agriculture, industry, and mineralogy~\cite{klein2007manual}. Recent studies have revealed the CSO as 
a spin-chain compound featuring fractional spinon excitations~\cite{Mourigal2013}, but the comprehensive 
phase diagram of this ubiquitous material remains unresolved. Here we establish its field-temperature phase 
--- through magnetocaloric and nuclear magnetic resonance (NMR) measurements down to about 20~mK ---
and reveal the existence of a low-temperature spin superfluid transition through 3D magnon BEC 
(Fig.~\ref{Fig1}\textbf{a}). This is in analogy to helium-4 phase diagram (Fig.~\ref{Fig1}\textbf{b}), 
whose low superfluid transition temperature renders helium-4 a unique quantum coolant. Here the
field suppression of the BEC transition temperature $T_c$ reveals a QCP at moderate 
fields --- a feature absent in helium-4 phase diagram. Besides the spin superfluid phase, 
we identify other regimes including a spin-chain Luttinger liquid and two distinct, 
cone-like quantum critical (QC) regimes. They contain finite-temperature 1D (Fermi gas) and 3D (Bose gas) 
QC states, respectively, separated by a dimensional crossover.

% ====== Fig. 2  ====== %
\begin{figure*}[htp]
\includegraphics[width=1\linewidth]{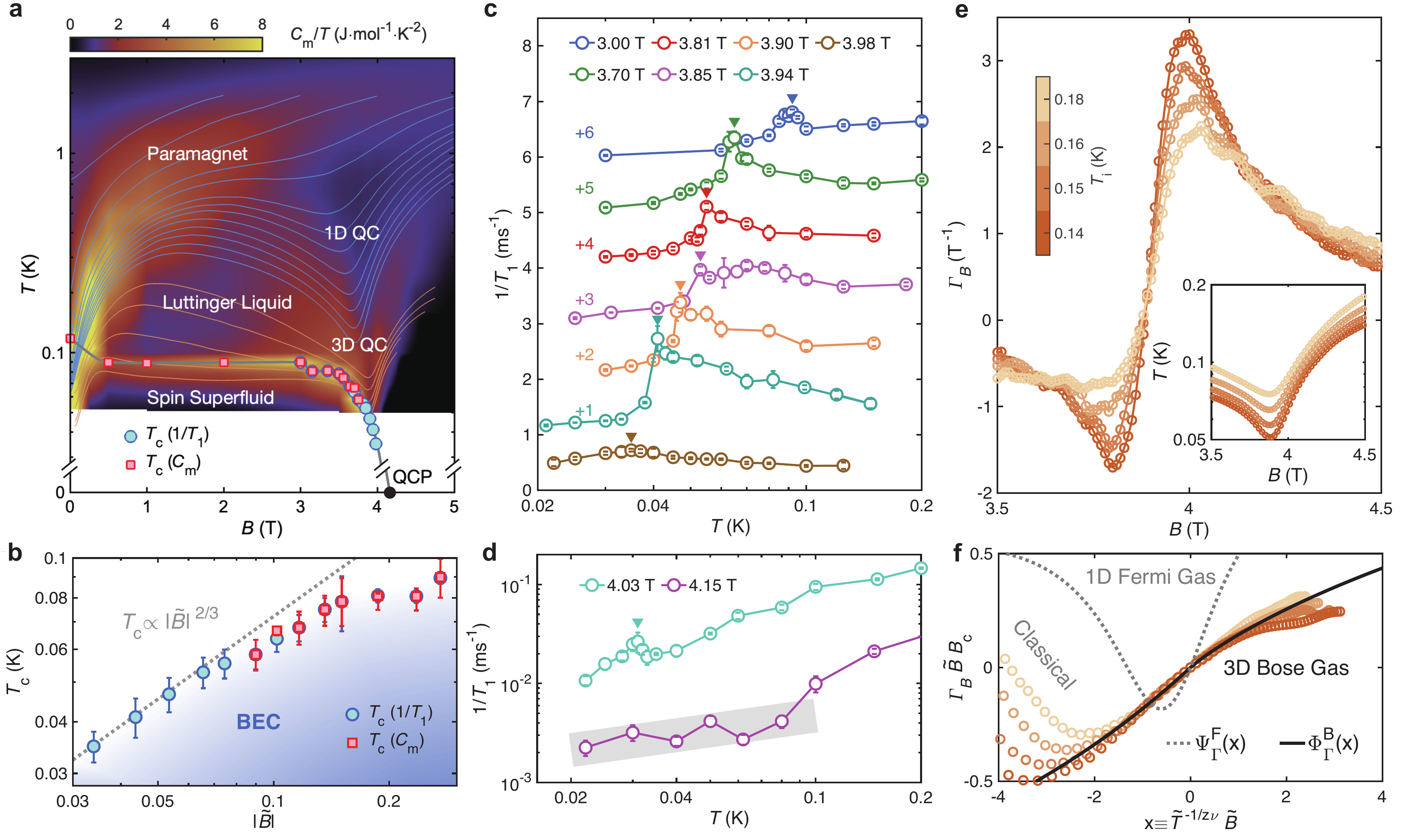}
\caption{\textbf{Field-temperature phase diagram of CSO and universal scaling behaviors.} 
\textbf{a} The color map represents the specific heat $C_m/T$, and the blue and yellow lines depict the 
measured isentropic lines in adiabatic demagnetization processes. The superfluid transition temperatures $T_c$ 
for magnon BEC, as determined from the spin-lattice relaxation rate ($1/T_1$) and specific heat 
data, are indicated by solid circles and squares, respectively. The gray line outlines the 
phase boundary of the spin superfluid phase, terminating at the BEC QCP at $B_c$ (black dot). 
\textbf{b} The transition temperatures follow a universal scaling $T_c\propto |\widetilde{B}|^{z/d}$, 
with $z=2$ and $d=3$ for the BEC QCP, where the reduced field reads $\widetilde{B} \equiv (B-B_c)/B_c$ 
(with $B_c \simeq 4.12$~T) . 
\textbf{c}, \textbf{d} illustrate the $1/T_1$ results measured by NMR under various fields, where
the superfluid transition temperatures are indicated. At $B = 4.15$~T, 
the $1/T_1$ shows no discernible peak down to 20~mK. The shaded region serves as a visual guide.
\textbf{e} Gr\"uneisen ratio $\Gamma_B$ derived from the low-temperature isentropic lines of inset. 
The colorbar represents the various initial temperatures $T_i$ at $B_i=4.5$~T.
\textbf{f} Data collapse of the Gr\"uneisen ratio from \textbf{e}, demonstrating scaling behavior 
consistent with the universality class of 3D BEC QCP. The black solid curve shows the scaling 
function $\Phi_{\Gamma}^{\rm B}(x)$ with $x\equiv \widetilde{B}\widetilde{T}^{-1/z\nu}$ for critical 3D 
Bose gas, while the gray dashed curve denotes $\Psi_{\Gamma}^{\rm F}(x)$ for 1D Fermi gas. Non-universal 
deviations, labeled as ``Classical", indicate the influence from finite-temperature superfluid transitions 
on the left side of the BEC QCP. 
} 
\label{Fig2}
\end{figure*}

The field-induced BEC QCP in CSO simultaneously overcomes 
both material-specific limitations and the ordering temperature ($T_c$) constraints, 
establishing CSO as an exceptional quantum coolant. 
While the initial demonstration of adiabatic demagnetization cooling with CSO reached 80~mK in 
1939~\cite{Ashmead1939Nature}, our work achieves a record low temperature of 12.8~mK, 
establishing a new benchmark for quantum materials refrigeration~\cite{Wolf2011,Lang2012,
Tokiwa2016,Xiang2017a,Xiang2024Nature}. The key difference involves applying moderate magnetic 
fields ($B > B_c \simeq 4.12$~T) to harness the universal quantum cooling emerging from the BEC 
QCP and its dimensional crossover. Relatedly, we observe very fast thermal relaxation near the QCP
--- significantly surpassing conventional paramagnetic salts --- due to the strong spin-phonon coupling. 
Our findings establish magnon condensation QCPs in quantum magnets as a new route for 
efficient millikelvin refrigeration.

\bigskip
\noindent
{\textbf{Thermodynamics measurements}}\\
Renowned for its easy accessibility and synthesis, CSO has played an active role in the studies of 
crystallography and magnetism: it was the first single crystal imaged via X-ray diffraction~\cite{Laue1913} 
and the first magnetic solid compound to exhibit NMR~\cite{Bloembergen1950NMR}. 
The low-temperature magnetothermal properties have been extensively characterized experimentally 
through pioneering works~\cite{Ashmead1939Nature,Wittekoek1964TwoMag,Giauque1967}. However, 
a comprehensive theoretical understanding of these properties has remained elusive for decades.

We synthesize centimeter-sized CSO single crystals through controlled cooling of supersaturated 
solutions (Supplementary Note~1). The magnetic structure (Fig.~\ref{Fig1}\textbf{e}) reveals two distinct copper sites: 
Cu-I (gray) forming antiferromagnetic Heisenberg $S=1/2$ chains along the $a$-axis, and Cu-II (green) 
hosting weakly interacting $S=1/2$ moments, as confirmed by density functional theory calculations 
(Supplementary Note~2).
 
The strong intrachain versus weak interchain couplings establish two distinct energy scales, making CSO an 
ideal system for studying spin Luttinger liquid physics, as demonstrated by neutron scattering observation of fractional
spinon excitations~\cite{Mourigal2013}. However, there exists pronounced specific heat anomaly near 100~mK --- 
first observed in Ref.~\cite{Ashmead1939Nature} and confirmed in our measurements (Supplementary Note~3). 
While traditionally attributed to 3D coupling effects~\cite{Tol1973HighField,Klaassen1977Order}, 
the nature of such a transition and its universality class have remained unresolved.

In Fig.~\ref{Fig2}\textbf{a}, we present the experimental phase diagram down to 20~mK, derived from magnetic specific 
heat ($C_m$), spin-lattice relaxation ratio ($1/T_1$) and magnetocaloric effect (MCE) measurements. The colormap of $C_m/T$ reveals enhanced 
thermal fluctuations at low temperatures across two distinct regions: one proximal to the QCP and the other adjacent to 
the zero field. The fan-shaped QC regimes arises from the QCP of correlated Cu-I spins; while in the low-field region, 
there are prominent specific heat contribution by paramagnetic Cu-II spins. Moreover, a distinct finite-temperature 
transition line is observed in the contour plot, which is a $\lambda$-type specific heat anomaly at about 100~mK 
(see Supplementary Note~3). The MCE measurements link the thermodynamic signatures to the QC behaviors of the system, 
and the isentropic lines obtained from adiabatic demagnetization measurements exhibit sharp minima near the QCP, 
where the finite-temperature transition line terminates. 

% ====== Fig. 3  ====== %
\begin{figure*}[htp]
	\includegraphics[width=1\linewidth]{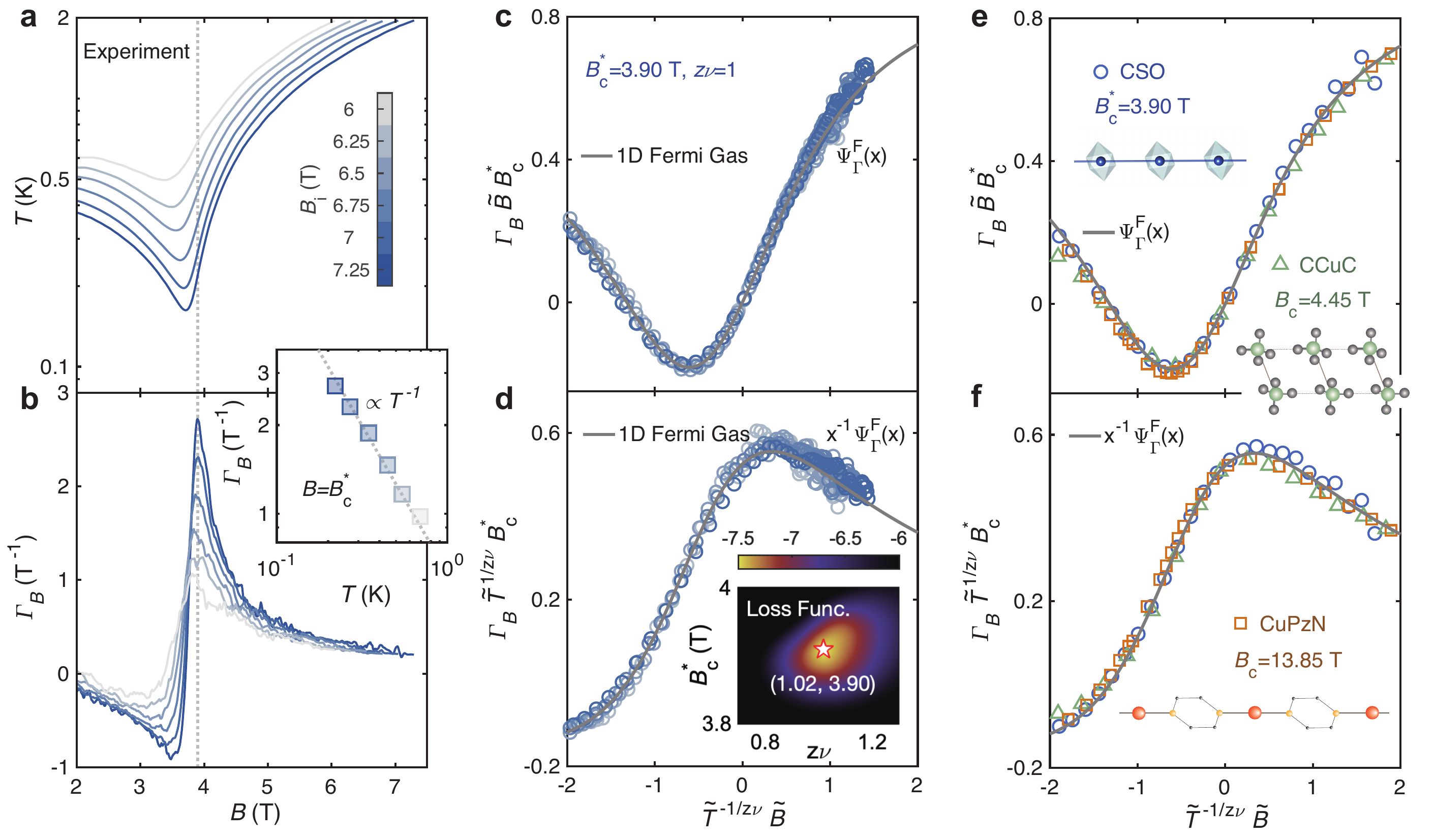}
	\caption{\textbf{Universal cooling in the 1D QC regime.}
		\textbf{a} Experimental measurements of isentropic lines by adiabatic demagnetization from about 2~K and
		various initial fields. The darker line represents the lower entropy, and the gray dashed line marks the 
		pseudo critical field $B_c^*\simeq 3.9$~T in finite-temperature perspective. 
		\textbf{b} Magnetic Gr\"uneisen ratio $\Gamma_B \equiv \frac{1}{T}\left(\frac{\partial T}{\partial B}\right)_S$, 
		extracted from \textbf{a}, shows peaks near $B_c^*$ (dashed line). Inset illustrates the divergence of the 
		Gr\"uneisen ratio $\Gamma_B(B_c^*,T)$, following a universal scaling $T^{-1/{z\nu}}$ with $z\nu=1$ for 1D QC. 
		\textbf{c,d} Two kinds of data collapse of Gr\"uneisen ratio. The gray lines represent the scaling function 
		$\Psi_{\Gamma}^{\rm F}(x)$ in \textbf{c} and $x^{-1}\Psi_{\Gamma}^{\rm F}(x)$ in \textbf{d} with 
		$x\equiv \widetilde{B}\widetilde{T}^{-1/z\nu}$ given by 1D Fermi gas. 
		Inset illustrates the loss function, ${L}(z\nu,B_c^*)\equiv \ln{\frac{1}{N} \sum_i \sigma^2(x_i)}$, 
		which quantifies the quality of data collapse. The pentagram marks the minimum of the loss function 
		($z\nu \simeq 1.02$ and $B_c^* \simeq 3.9 \,{\rm T}$, given an optimal $g\simeq 1.82$). 
		\textbf{e,f} The data collapsing results of different magnetic materials: CSO, \CPN and \CCC (CCuC), 
		with different critical fields and crystal structures. \CPN is a spin-chain material with $B_c\simeq 13.85$~T
		\cite{Breunig2017SA}, while CCuC is a spin-ladder compound with $B_c\simeq 4.45$~T~\cite{CCCMCE2014}. 
	} 
	\label{Fig3}
\end{figure*}

\bigskip
\noindent
\textbf{NMR and MCE measurements of 3D BEC QCP} \\
To explore the 3D phase transition boundary and the universality class of the QCP, we conducted $^1$H 
NMR measurements on a CSO single crystal. Above the transition temperature $T_c$, only four NMR lines 
are observed, which stem from the H(6-10) sites adjacent to Cu-I ions (fig.~\ref{Fig_SM_NMR_spectra}). 
In Figs.~\ref{Fig2}\textbf{c} and \textbf{d}, we present the spin-lattice 
relaxation rate 1/$T_1$ measured at the $l_2$ line for varying magnetic fields (fig.~\ref{Fig_SM_NMR_all}). 
For 3.00~T $\leq B \leq$ 3.98~T, the obtained 1/$T_1$ exhibits a clear peak at $T_c$ , 
a typical characteristic of the magnetic phase transitions~\cite{BCVO2015}. At 4.03 T, spin fluctuations weaken 
with decreasing temperature, reducing $1/T_1$  until a small peak appears at $T_c$. 
In contrast, at 4.15 T, $1/T_1$  decreases monotonically to the lowest temperatures, showing no magnetic ordering. 
These characteristics are consistently reproduced at the $l_3$  line (fig.~\ref{Fig_SM_P2_T1}).

Using the extracted 3D transition temperatures $T_c$ from NMR and specific heat data at each field, we map out the phase 
boundary (gray line in Fig.~\ref{Fig2}\textbf{a}). In Fig.~\ref{Fig2}\textbf{b}, the $T_c$ dataset includes specific 
heat peaks, which align with NMR results. The key evidence supporting the identification of a BEC QCP in 
CSO comes from the $1/T_1$ peaks extending down to about $35$~mK. We find the 3D transition line follows 
a universal scaling law near the QCP, i.e., $T_c \propto |\widetilde{B}|^{z/d}$, where $\widetilde{B} 
\equiv (B - B_c)/B_c$ is the reduced field. The estimated critical field is $B_c \simeq 4.12$~T, above which the spins become fully 
polarized. The dynamical exponent $z = 2$ and the spatial dimension $d = 3$, thus $z/d=2/3$, place the 
system within the universality class of a 3D BEC QCP. The observed 2/3-scaling law is derived from the 
Hartree-Fock-Popov description~\cite{Oshikawa2000BEC}, which remains valid at relatively low boson 
densities consistent with prior studies~\cite{Mladen2017}. 
It is noteworthy that an early study determined the critical exponent for the finite-temperature transition at $T_c$ to be 
$\beta \simeq 0.34 \pm 0.04$~\cite{Klaassen1977Order}. In retrospect, this value well supports the spin superfluid 
phase transition ($\beta \simeq 0.349$ for 3D XY universality class). Here we show such finite-temperature 
transitions end up at a BEC QCP as field reaches the critical value, as evidenced by our NMR measurements. 

Besides NMR, we also conduct low-temperature MCE measurement down to the 3D QC regime. The universally 
diverging Gr\"uneisen ratio can be used to sensitively locate the QCP~\cite{Zhu2003,Zhitomirsky2004,Garst2005,
Wolf2011,Wolf2016,Xiang2017a,Liu2021PRR,Liu2022,Xiang2023,Xiang2024Nature,li-2024-nc}. With a dilution 
refrigerator insert, we start from a low initial temperature below 200~mK, and attain a minimum temperature of 
about 50~mK near the QCP. Such a significant cooling effect is characterized by the diverging magnetic Grüneisen 
ratio $\Gamma_B \equiv \frac{1}{T} \left( \frac{\partial T}{\partial B} \right)_S$, as shown in Fig.~\ref{Fig2}\textbf{e}. 
A characteristic peak-dip structure emerges in $\Gamma_B$, with both features sharpening dramatically upon cooling --- 
direct evidence of QCP. The observed scaling collapse of thermal data onto a universal function provides unambiguous 
experimental verification of magnon BEC universality class.

To be specific, the 3D QC regime exhibits universal behaviors described by critical 3D Bose gas
(see Fig.~\ref{Fig1}\textbf{c}, and also Methods), according to the universal scaling form 
$$\Gamma_B(B,T) = \frac{1}{\widetilde{B} B_c} \, 
\Phi_{\Gamma}^{\rm B}(\widetilde{B}\widetilde{T}^{-1/z\nu}),$$ where $\widetilde{T} \equiv T/T_0$ with 
$T_0 = g \mu_B B_c/k_B$ is the rescaled temperature, {$z=2$ and $\nu=1/2$ are critical exponents of the BEC QCP}. 
In this manner, the experimental data collapse 
onto a universal curve $\Phi_\Gamma^{\rm B}$ (Fig.~\ref{Fig2}\textbf{f}), in excellent agreement with 
the scaling function of 3D critical Bose gas, with $x\equiv \widetilde{B}
\widetilde{T}^{-1/z\nu}$. In Fig.~\ref{Fig2}\textbf{f}, we also plot the $\Psi_{\Gamma}^{\rm F}(x)$ of 1D 
Fermi gas, which is distinct from $\Phi_{\Gamma}^{\rm B}(x)$ of 3D Bose gas. Therefore, the excellent 
data collapse  provides compelling evidence for a 3D BEC QCP. On the other hand, for $x<0$ the system 
can undergo a finite-temperature superfluid transition where the 3D Bose gas description becomes 
inadequate. This breakdown is evidenced by the relatively poor data collapse observed in the ``Classical'' 
regime of Fig.~\ref{Fig2}\textbf{f}.

% ====== Fig. 4  ====== %
\begin{figure*}[htp]
\includegraphics[width=1\linewidth]{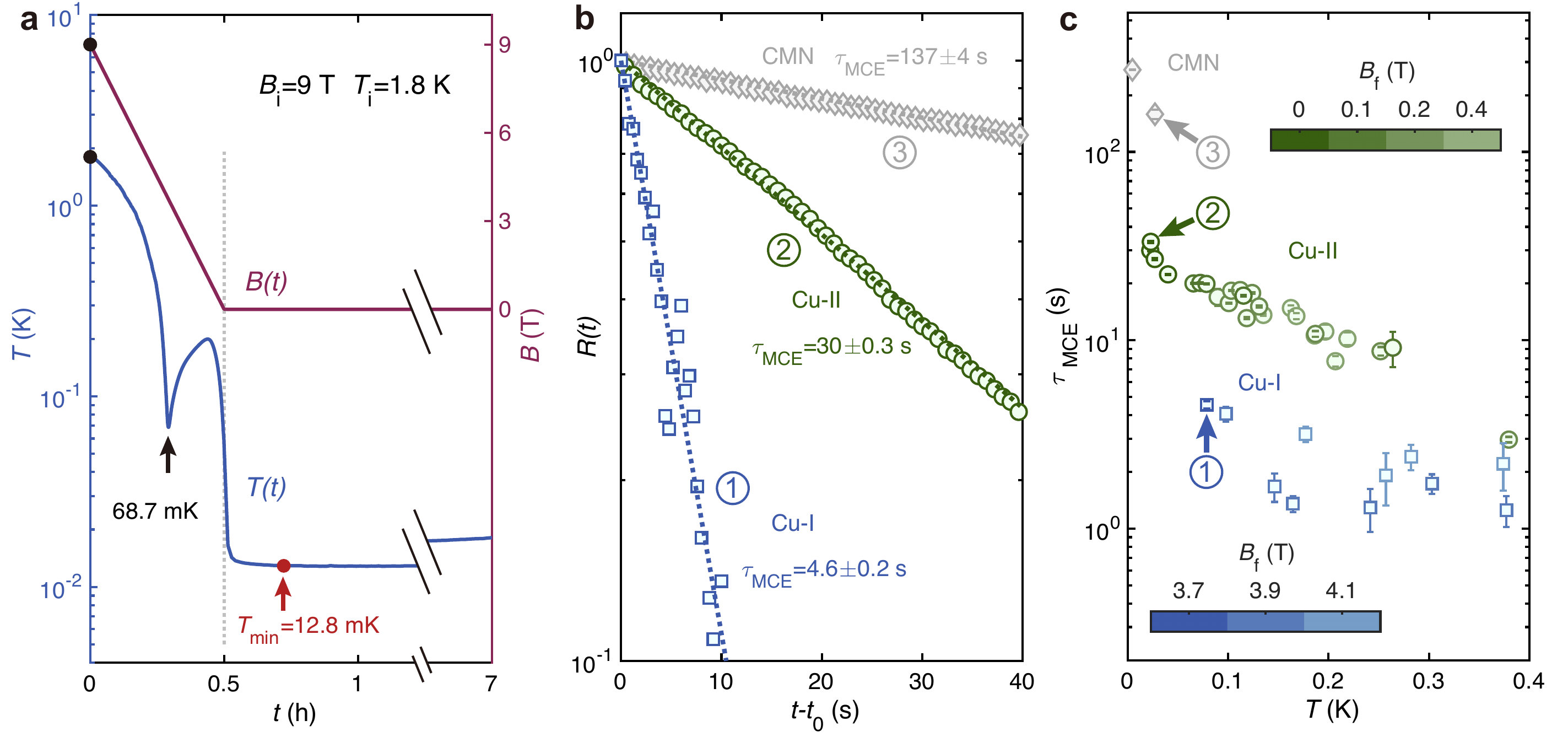}
\caption{\textbf{Adiabatic demagnetization cooling and magnetocaloric relaxation time.}
\textbf{a} The adiabatic cooling measurements. The black 
dot marks the initial condition at 9~T and 1.8~K, while the red dot indicates the lowest achieved temperature 
of 12.8~mK at zero field. The system maintains sub-20 mK temperatures for more than 7 hours, showing
very long hold time. The vertical dashed line indicates zero magnetic field.
\textbf{b} The thermal relaxation processes at various magnetic fields. Three representative cases, labeled 1-3, 
correspond to Cu-I, Cu-II, and the typical paramagnetic hydrate salt \CMN (CMN), respectively. The measured 
relaxation curves display exponential decay, allowing extraction of the thermal relaxation times $\tau_{\rm MCE}$ via 
fitting.
\textbf{c} The magnetocaloric relaxation time $\tau_{\rm MCE}$ for different fields and temperatures. $\tau_{\rm MCE}$ 
measured near the QCP (blue square) is governed by coupled Cu-I, and the low-field measurement (green circle) 
is governed by decoupled Cu-II. The gray diamonds illustrate the $\tau_{\rm MCE}$ of CMN. Circles 1-3 mark 
three representative cases with their relaxation curves in \textbf{b}.
} 
\label{Fig4}
\end{figure*}

\bigskip
\noindent
\textbf{1D QC regime via dimensional crossover} \\ 
As the temperature rises above approximately 200 mK, the 3D interchain coupling becomes negligible, 
leading to a dimensional crossover (Fig.~\ref{Fig1}\textbf{a}). In this regime, a 1D QC behavior emerges, 
described by 1D Fermi gas theory (Fig.~\ref{Fig1}\textbf{d}, and see Methods). 
Figure~\ref{Fig3}\textbf{a} shows the obtained isentropic lines starting from a higher initial temperature 
of about 2~K,  which reach minimum temperatures within the 1D QC regime. In Fig.~\ref{Fig3}\textbf{b} 
the corresponding Gr\"uneisen ratio results are shown, where a diverging $\Gamma_B \propto T^{-1/z\nu} 
=T^{-1}$ is observed. Through minimization of the loss function $L(z\nu,B_c^*)$ (see inset in 
Fig.~\ref{Fig3}\textbf{d}), we estimate pseudo-critical field $B_c^*\simeq 3.9$~T and critical exponent $z\nu 
\simeq 1.02$. The latter is in excellent agreement with the expected product of critical exponents $z=2$ and $\nu=1/2$ for the 1D QC. 
Moreover, data collapse analysis yields a universal scaling form $$\Gamma_B(B,T) 
= \frac{1}{\widetilde{B}B_c^*} \Psi_{\Gamma}^{\rm F}(\widetilde{B} \widetilde{T}^{-1/z\nu}),$$ as shown in 
Fig.~\ref{Fig3}\textbf{c} and its variation in Fig.~\ref{Fig3}\textbf{d}. Here for 1D QC the reduced field is defined 
as $\widetilde{B} \equiv ({B-B_c^*})/{B_c^*}$, and $\widetilde{T} \equiv T/T_0$ with $T_0 = g\mu_B B_c^*/k_B$. 
The experimental data in Fig.~\ref{Fig3}\textbf{a,b} show excellent collapse onto the scaling function 
$\Psi_{\Gamma}^{\rm F}(x)$ derived from 1D critical Fermi gas (see Methods). 

We observe universal scaling of the Gr\"uneisen ratio that is independent of microscopic particulars. This scaling 
behavior persists across chemically diverse compounds, suggesting a fundamentally underlying principle. In 
Fig.~\ref{Fig3}\textbf{e,f}, we show three different materials --- the spin-chain compound \CPN with $B_c = 13.85 \, 
\text{T}$~\cite{Breunig2017SA}, the spin-ladder compound \CCC (abbreviated CCuC) with $B_c = 4.45 \, \text{T}$~\cite{CCCMCE2014}, 
as well as CSO --- exhibit Gr\"uneisen ratios that collapse onto the same universal scaling function of the 1D critical 
Fermi gas, i.e., $\Psi_{\Gamma}^{\rm F}(x)$. Despite differences in coupling strengths, critical fields, and even lattice 
geometries, the shared universality class ensures identical scaling behavior in $\Gamma_B$. 
These results demonstrate that the universality classes of quantum critical behaviors can be precisely measured using MCE. 
\\
	
\bigskip
\noindent
{\textbf{Magnetocaloric relaxation}} \\
Our adiabatic demagnetization measurements on CSO (Fig.~\ref{Fig4}\textbf{a}) demonstrate cooling to 
68.7~mK near the QCP and a record-low 12.8~mK at zero field. This unprecedented performance stems 
from self-cascaded cooling mechanism: universal QCP cooling from Cu-I ions (68.7 mK) combined with paramagnetic 
cooling from nearly free Cu-II ions (12.8~mK). The synergistic MCE between these subsystems produces 
a collective effect exceeding simple summation, establishing a new benchmark for quantum magnetic 
refrigeration.

The BEC QCP enables both ultralow temperature cooling and rapid thermalization. Practical measurements 
show thermal lag --- temperature continues dropping after reaching zero field (Fig.~\ref{Fig4}\textbf{a}), 
gradually approaching $T_{\rm min}$, the equilibrium (lowest) temperature under the fixed field~\cite{Isono2018}. 
This reflects the magnetocaloric thermalization process --- spins cool via demagnetization while the lattice 
remains warmer, with equilibrium established on finite relaxation timescales $\tau_{\rm MCE}$.

To study the thermal relaxation between the ``cold'' spin and the ``warm'' lattice, in the adiabatic 
demagnetization process, we pause the field at a fixed value and measure the temperature variations 
in the relaxation process (see Methods and Supplementary Note~5). Figure~\ref{Fig4}\textbf{b} 
illustrates the temperature variation ratio $R(t) \equiv \frac{T(t)-T_{\rm min}}{T(t_0)-T_{\rm min}}$. 
We find $R(t)$ falls into an exponential scaling with time, i.e., $R(t) = e^{-(t-t_0)/\tau_{\rm MCE}}$, 
with $t_0$ the time we pause the field and $\tau_{\rm MCE}$ the magnetocaloric relaxation time. 
As shown in Fig.~\ref{Fig4}\textbf{b}, in the 3D QC regime, the cooling effect is mainly contributed by 
Cu-I ions, whose magnetocaloric relaxation time is found to be very short, i.e., $\tau_{\rm MCE} \simeq 4.6$~s. 
Under zero field, on the other hand, the relaxation dynamics are mainly related to the nearly free Cu-II spins, 
exhibiting a significantly prolonged $\tau_{\rm MCE} \simeq 30$~s --- much longer than those near 
the BEC QCP and at equivalent temperatures. 

Comparing CSO with paramagnetic hydrate salt \CMN (CMN) at similar temperatures (Fig.~\ref{Fig4}\textbf{b}), 
we find $\tau_{\rm MCE} \simeq 30$~s for CSO at 23 mK --- its longest relaxation time but still an order 
of magnitude shorter than $\tau_{\rm MCE} \simeq 137$~s of CMN at slightly higher temperature. The 
fitted $\tau_{\rm MCE}$ values are summarized in Fig.~\ref{Fig4}\textbf{c}. The enhanced thermalization 
of CSO stems from its coupled Cu-I spin system, where low-energy magnons efficiently scatter phonons, 
enabling rapid magnetocaloric relaxation even at ultralow-temperatures. Notably, Fig.~\ref{Fig4}\textbf{c} shows 
that Cu-I and Cu-II exhibit distinct $\tau_{\rm MCE}$ despite equivalent temperatures, indicating that 
phonon conductivity itself cannot account for the observed cooling lag. The dynamics are instead governed by 
spin-phonon coupling, which constitutes the discriminating factor for both CSO subsystems and CMN (see Methods).
\\

\noindent{\textbf{Discussion}}\\  
Despite being a very common compound, the quantum magnet CSO exhibits remarkably rich 
physics. Our ultralow-temperature MCE and NMR measurements identify a magnetic-field-driven 
BEC QCP. Our exact theoretical solutions quantitatively account for the universal MCE scaling 
observed in real materials over a wide temperature range. 
Historically, Giauque \textit{et al.} conducted exhaustive studies on the low-temperature properties 
of CSO~\cite{Geballe1952Thermo,Giauque1967,Giauque1968II,Giauque1968III,Giauque1968IV,
Giauque1970V,Giauque1975VI}, and concluded it was unsuitable as a thermometric reference 
material~\cite{Giauque1970V}, due to its complex cooling behavior. In contrast, here we reveal 
universal cooling behaviors emerges near the BEC QCP --- a material-independent phenomenon 
characterized by the scaling function, establishing CSO as an excellent thermometric standard.

In addition, fast magnetocaloric relaxation rate near the BEC QCP renders CSO exceptional cooling 
efficiency. In conventional paramagnetic salts, weak spin-phonon coupling leads to rather long magnetocaloric
relaxation times ($\tau_{\rm MCE} > 10^2$~s at millikelvin temperatures), creating a thermal bottleneck 
that severely limits cooling power (e.g., $\lesssim 1~\mu$W at 50~mK)~\cite{Shirron2007}. 
Remarkably, this limitation can be overcome in CSO as the entropies are 
carried by collective excitations rather than local spins. The critical Bose or Fermi gas emerging 
near the QCP can efficiently scatter low-energy phonons, enabling fast magnetocaloric relaxation 
on the order of seconds. The simultaneous achievement of low cooling temperatures, high 
magnetic entropy density, and rapid thermal relaxation rates provides crucial insights for resolving 
the persistent challenges in ADR. 

Another point we would like to emphasize involves self-cascaded cooling between spin subsystems 
with hierarchical coupling strengths. The realization of nearly 10~mK temperature with CSO is enabled 
by its complementary dual Cu-I and Cu-II magnetic systems. At the critical field $B_c \simeq 4.12$~T, 
the Cu-I spins first undergo universal cooling; as the field further decreases, the nearly free Cu-II spins 
activate a secondary cooling stage near zero field. This quantum handoff creates a continuous cooling 
trajectory unmatched in conventional refrigerants. 

From a practical perspective, millikelvin refrigeration serves as the backbone for multiple cutting-edge 
technologies, including quantum computing, space applications, and advanced electronics, etc
\cite{Shirron2014,Jahromi2019nasa}. Liquid helium has been a key quantum coolant for cryogenic 
refrigeration due to its strong zero-point fluctuations and low-temperature superfluid transition
\cite{Frank2007ThermalC}. However, there is a fundamentally limitation with helium-4 --- the lowest 
cooling temperature is around 1~K in practice. Reaching the millikelvin regime currently relies on 
scarce and costly helium-3 isotope~\cite{Cho2009Science,Kramer2019Helium}, creating a critical 
demand for alternative refrigerants. Here we show CSO constitutes a remarkably cost-effective and 
efficient millikelvin coolant. It thus demonstrates how magnon BEC --- a fundamental quantum phenomenon
 --- can be practically utilized for extreme cryogenic refrigeration.

\bigskip
\noindent
{\textbf{Methods}} \\
\textit{3D Bose gas theory. ---}
We consider the Bose field theory 
\begin{equation}
\mathcal{H}_{\rm B} = \int {\rm d}x \left( \frac{1}{2m} |\nabla \psi_{\rm B}|^2 + \bar{B} |\psi_{\rm B}|^2 
	+ \frac{u_0}{2} |\psi_{\rm B}|^4 \right), 
\notag
\end{equation}
to describe the universal properties of CSO near the BEC QCP at $\bar{B}=0$. $\bar{B}$ represents 
the chemical potential of Bose gas and corresponds to the magnetic field $\bar{B} \equiv B-B_c$ in spin 
system. As $d+z=3+2>4$, it lies above the upper-critical dimension, and the interaction is irrelevant 
due to the negative scaling dimension, ${\dim} \left[u_0\right]=-1<0$. Consequently, mean-field 
theory accurately describes the system near the QCP~\cite{sachdev2015}. 
To minimize the free energy, the boson density is 
\begin{equation}
\braket{\psi_{\rm B}^{\dagger} \psi_{\rm B}}=
\begin{cases} 
	& -\bar{B}/u_0 + \cdots, \quad \bar{B} < 0, \\ 
	& \quad \quad 0, \quad \quad \quad \,\, \quad \bar{B} > 0. 
\end{cases}  
\notag
\end{equation}

\begin{figure}[htp]
\includegraphics[width=0.9\linewidth]{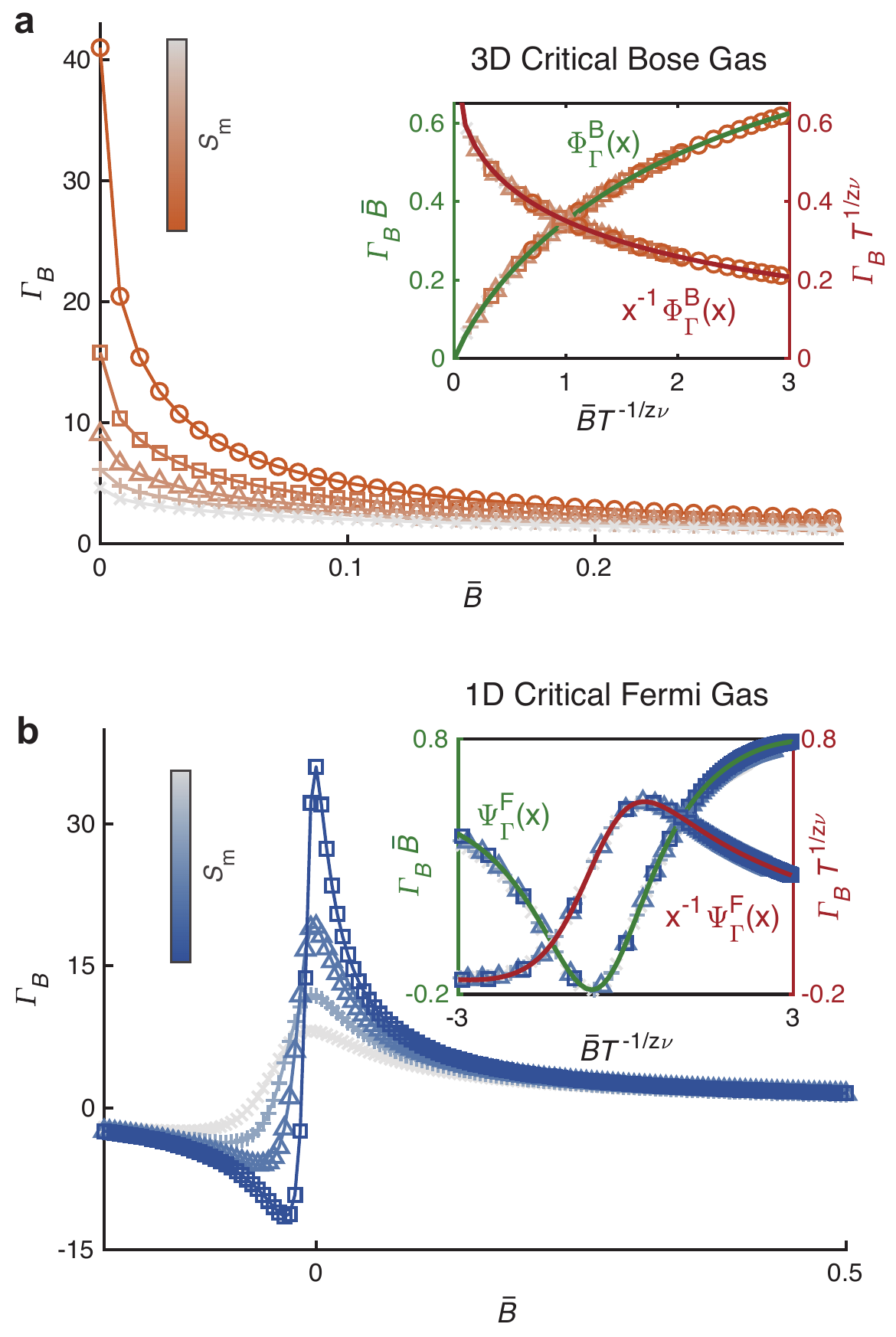}
\caption{The Gr\"uneisen ratio $\Gamma_B$ of \textbf{a} 3D critical Bose gas and \textbf{b} 
1D critical Fermi gas. The two insets show the scaling functions $\Phi_{\Gamma}^B(x)$ 
and $\Psi_{\Gamma}^F(x)$ obtained via data collapsing, where $x\equiv \bar{B} 
{T}^{-1/z\nu}$. Taking the 3D Bose gas as an example, the universal 
functions are obtained following $\Gamma_B = {\bar{B}^{-1}} \, 
\Phi_{\Gamma}^{\rm B}(\bar{B}{T}^{-1/z\nu}) = {T^{-1/z\nu}} \, 
x^{-1} \Phi_{\Gamma}^{\rm B}(\bar{B}{T}^{-1/z\nu})$.
} 
\label{Figm1}
\end{figure}

As for $\bar{B}<0$, the critical temperature $T_c$ of superfluid transition is determined 
by the two-body interaction $u_0$ and field $\bar{B}$ with a universal scaling, 
\begin{equation}
T_c = \frac{2\pi}{m} \left( \frac{|\bar{B}|}{2u_0\zeta(3/2)} \right)^{2/3} 
\propto |\bar{B}|^{2/3},
\end{equation}
which has been observed experimentally in magnon BEC transition in quantum magnets
\cite{Batista2014RMP}.

As for $\bar{B}>0$ regime, the interaction $u_0$ is irrelevant and safely ignored. Thus, 
the critical Bose gas is described by a free field theory, 
\begin{equation}
\begin{split}
f_{\rm B} & = k_B T \int \frac{{\rm d}^3 k}{(2\pi)^3} \log{\left[1-e^{(-\bar{B}-\frac{k^2}{2m})/T} \right]} \\
	      & = C_{\rm B} T^{5/2} \Lambda_f^{\rm B}\left(\frac{\bar{B}}{T}\right), 
\end{split}
\label{Eq:FreeEnergy_3DBoson}
\end{equation}
where $\Lambda_{f}^{\rm B}(x) = - \left(\frac{1}{2\pi}\right)^{3/2} {\rm Li}_{5/2} \left(e^{-x}\right)$ is the 
scaling function of free energy, ${\rm Li}_n(z)\equiv \sum_{k=1}^{\infty}z^k/k^n$ is the polylogarithm 
function, and $C_{\rm B}=m^{3/2} k_B$ is a specific constant. We can then derive thermodynamic 
properties from the free energy, which exhibit universal behaviors and quantum universal scaling. 
For example, the magnetic Gr\"uneisen ratio, which corresponds to second derivative of free energy, 
can be expressed as, 
\begin{equation}
\Gamma_{B} = T^{-1} \Phi_{\Gamma}^{\rm B}\left(\frac{\bar{B}}{T}\right), 
\notag
\end{equation}
where $\Phi_{\Gamma}^{\rm B}(x)$ is the scaling function of 3D critical Bose gas. At the QCP, the 
Gr\"uneisen ratio diverges following $\Gamma_B(B_c,T)=\Phi_{\Gamma}^{\rm B}(0) T^{-1} \propto 
T^{-1}$ ($\bar{B}=0$).

\textit{1D Fermi gas theory. ---}
With increasing temperature, the system undergoes a dimensional crossover from 3D to 1D quantum 
critical behavior as interchain couplings become irrelevant. To describe the latter, we consider a 1D 
quantum XX model, $H_{\rm XX} = J_{x} \sum_{i} ( S_i^x S_{i+1}^x + S_i^y S_{i+1}^y ) - B \sum_i S_i^z$. 
Through the Jordan-Winger transformation, the Hamiltonian can be represented (in the momentum space) 
as a quadratic, non-interaction fermion chain $H_{\rm XX} = \sum_k \varepsilon_k c_k^{\dagger}c_k$, 
with $\varepsilon_k  = J_{x} \cos{(ka)} + B_z$. At the 
critical point $B_c = J_{x}$, the low-energy dispersion is quadratic, $\varepsilon_k \propto k^2$, 
with a dynamic exponent $z=2$. In the continuum limit, the Hamiltonian can be expressed as 
\begin{equation}
    \mathcal{H}_{\rm F} = \int {\rm d}x \left( \frac{1}{2m} |\nabla \psi_{\rm F}|^2 + 
	\bar{B} |\psi_{\rm F}|^2  \right), 
    \label{Ham_Fermion}
    \notag
\end{equation}
where $\psi_{\rm F}$ is the Fermi field operator, $m=1/(J_{x}a^2)$ is the fermion mass.
$\bar{B}$ represents the chemical potential, where $\bar{B} = B-B_c$ also corresponds 
to the magnetic field in the spin system. Using the scaling dimension, 
${\rm dim}[x] = -1$ and ${\rm dim}[\tau] = -2$, 
we find that the relevant perturbation $\bar{B}$ has a scaling dimension of ${\rm dim}[\bar{B}] = 2$, 
leading to a critical exponent $\nu=1/2$. Given the Fermi statistics, the free energy can be 
expressed as 
\begin{equation}
\label{EqM:FF}
\begin{split}
	f_{\rm F} & = - k_B T \int \frac{{\rm d} k}{2\pi} \log\left[{1 + e^{(-\bar{B}-
	\frac{k^2}{2m})/T}}\right] \\ 
	& = C_{\rm F} T^{3/2} \Lambda_{f}^{\rm F}\left(\frac{\bar{B}}{T}\right), 
\end{split}
\end{equation}
where $\Lambda_{f}^{\rm F}(x) = \frac{1}{2\pi} \int {\rm d}y \log{(1+e^{-x-y^2})}$ 
is the scaling function of free energy and $C_{\rm F} = -({2m})^{1/2}k_B$ is a constant. 
From the free energy Eq.~(\ref{EqM:FF}), the magnetic Gr\"uneisen ratio reads, 
\begin{equation}
	\Gamma_{B} = T^{-1} \Psi_{\Gamma}^{\rm F}\left(\frac{\bar{B}}{T}\right), 
	\notag
\end{equation}
where $\Psi_{\Gamma}^{\rm F}(x)$ is the scaling function of 1D critical Fermi gas. 
Similarly, we also have $\Gamma_B(B_c^*,T) \propto T^{-1}$ at the QCP with $\bar{B}=0$.

\textit{NMR measurements.---}
The measurements are carried using a phase-coherent pulsed NMR spectrometer, on a single crystal 
of about $9 \times 4 \times 2$ mm$^3$ size. Below $T$ = $1.5$~K, measurements are performed 
by utilizing a dilution refrigerator. $^1$H-NMR spectra are acquired by adhering to the fast Fourier 
transforms of each spin-echo signal, while the spin-lattice relaxation rate $1/T_1$ is measured 
through the saturation-recovery method~\cite{Abraham1961}. The radio frequency coil is specially 
fabricated with pure copper wire covered by teflon to eliminate the proton NMR signal outside the 
sample. The external magnetic field is applied parallel to the $\gamma^\ast$ axis. More details
of the NMR measurements can be found in Supplementary Note~6. 
\\

\textit{MCE measurements.---}
The measurements are conducted through adiabatic demagnetization process, carried out using 
home-designed device~\cite{Xiang2023}. The measurements were implemented within a 
commercial Quantum Design Physical Property Measurement System (PPMS). The two-layer device features 
a straw-like structural design that optimizes thermal isolation. The outer layer contains the 
Gd$_3$Ga$_5$O$_{12}$, which serve as thermal guard against PPMS chamber heat leaks, while 
the inner layer contains $2.7$~g (about $1.21$~cm$^3$) CSO crystals, which functions as the primary 
cooling stage. A ruthenium oxide thermometer is embedded at the center of the sample, which has 
been calibrated down to ultralow temperature (see Supplementary Note~4). Copper wires, approximately 
$10$~cm in length and $100~\mu$m in diameter, are wound around the samples to enhance thermal 
exchange. To minimize thermal leakage between the sample and the outer layers, electrical signal 
connections are made using $25~\mu$m-diameter manganin wires. The measurement process begins 
in an environment with helium-4 exchange gas, allowing the sample temperature to be equal with the 
environment temperature of the chamber. Subsequently, a high vacuum state is achieved using the 
cryopump of the system to realize the adiabatic conditions. The demagnetization process is then 
conducted, with the field sweep rate set at $0.3$~T$\cdot$min$^{-1}$. 

The MCE measurements with initial temperature below about $200$~mK was conducted 
using a dilution refrigerator insert equipped with a commercial heat capacity option, whose 
built-in thermometer was calibrated under magnetic fields. In practice, a single-crystal sample 
of $7.8$~mg mass was used. During the demagnetization process, the magnetic field was 
varied at a rate of $0.03$-$0.12$~T$\cdot$min$^{-1}$.
\\
		
\textit{Thermal relaxation rate.---}
We obtain the magnetocaloric relaxation time $\tau_{\rm MCE}$ also with the home-designed double-layer device for MCE
measurements, and through a stepwise field-sweep protocol. Starting at about 2~K and 4-9~T field, 
we reduce the field in 0.1-0.2~T steps (0.9~T$\cdot$min$^{-1}$), pausing at key points including $B_c$ and 
zero field to record relaxation dynamics. The field-holding period extends up to 2000~s --- more 
than sufficient for complete thermal equilibration at $T_{\rm min}$. Parallel measurements on paramagnetic hydrate salt CMN 
($2.6$~g, $1.21$~cm$^3$, volume-matched to CSO) show $\tau_{\rm MCE} \approx 137$~s at 27~mK --- 
over 4 times longer than that of CSO at a similar temperature. The results indicate that the relaxation 
time for CMN is on the order of minutes, which is substantially longer than that of CSO. This stark 
difference persists across all measured temperature regimes (c.f, Fig.~\ref{Fig4}\textbf{c}).

Our experimental data (Fig.~\ref{Fig4}) strongly indicate that spin-phonon coupling --- rather than 
phonon thermal conductivity --- governs magnetocaloric relaxation dynamics. We substantiate this through 
relaxation rates. Assuming thermal conductivity of CSO $\kappa_{\rm CSO} \sim 10^{-4}$~W$\cdot$cm$^{-1}\cdot$K$^{-1}$ 
(estimated based on hydrate materials) at about 0.1~K~\cite{Frank2007ThermalC}, we proceed to analyze the thermal 
transport timescales. For a columnar-type sample, $K_{\rm CSO} = \kappa_{\rm CSO} \cdot A/L$, where 
$A\simeq 1$~cm$^{2}$ is the base area and $L\simeq 1$~cm is the length of the sample. The lattice heat 
capacity of the sample is $C_{\rm L} = c_{\rm ph} \cdot V \cdot \rho_{\rm CSO} / M_{\rm CSO}$, where 
$c_{\rm ph} \simeq 10^{-5}$~J$\cdot$mol$^{-1}\cdot$K$^{-1}$ represents the phonons contribution at about 0.1~K (see 
Supplementary Note~3), $V=A\cdot L \simeq 1$ cm$^{3}$ is the volume of the sample, 
$\rho_{\rm CSO} \simeq 2.23$~g$\cdot$cm$^{-3}$ is the density and $M_{\rm CSO} 
\simeq 249.7$~g$\cdot$mol$^{-1}$ is the molar mass of CSO. With this, we estimate 
the thermal transport time scale $\tau_{\rm L} \simeq C_{\rm L}/K_{\rm CSO} 
\sim 10^{-3}$~s, five orders of magnitude faster than the observed $\tau_{\rm MCE}\sim 10$~s 
at about 0.1~K. This clear discrepancy shows that the phonons equilibrate nearly instantaneously, 
while spin-phonon coupling governs the slow dynamics. This fundamental timescale separation 
explains why quantum critical systems outperform conventional paramagnetic refrigerants.

\bibliography{MCERef.bib} 
		
$\,$\\
\section*{Acknowledgments}
The authors (E.L., Y. Q., and W.L.) are indebted to Long Zhang for helpful discussions. 
We thank Heng Tu, Guochun Zhang for the preparation of the CMN crystals.

\paragraph*{Funding:}
This work was supported by the National Key R \& D Program of China grant number 
2024YFA1409200 and 2023YFA1406103, the National Natural Science Foundation of China 
(Grant Nos.~12222412, 12047503, 12074024, 52088101, 12374142, 12304170, 
12404180 and 12141002), Strategic Priority Research Program and {Scientific Instrument 
Developing Program} of CAS (Grant Nos.~XDB1270000, XDB1230100, ZDKYYQ20210003), CAS Project 
for Young Scientists in Basic Research (Grant No.~YSBR-057), and the Fundamental Research 
Funds for the Central Universities in China. We thank the HPC-ITP for the technical 
support and generous allocation of CPU time. {A portion of this work was carried out at the 
Synergetic Extreme Condition User Facility (SECUF, https://cstr.cn/31123.02.SECUF).}	

\paragraph*{Author contributions:}
W.L., J.X., and W.J. initiated this work, J.X., X.-Y.L., and P.S. designed and conducted 
the low-temperature MCE measurements, E.L., D.W.Q., W.L., Y.Q., and G.S. made the 
model calculations and theoretical analysis, X.T.H. and X.C. made the DFT calculations, 
C.S., X.W., and X-Y.L. prepared the single-crystal sample, S.Q., S.L., J.L., Y.J. and R.Z. conducted
the NMR measurements and analyzed the data. J.X., C.S., Y.Z., Q.Z., H.L., and W.J. performed the 
low-temperature specific heat measurements, J.X., E.L., X.-Y.L., W.J., and W.L. analyzed 
the experimental data, W.L. and G.S. supervised this project, E.L., J.X., W.L., and G.S. 
wrote the manuscript with contributions from all coauthors.

\paragraph*{Competing interests:}
The authors declare no competing interests. 
		
\paragraph*{Data and materials availability:}
The data that support the findings of this study are available at 
The code that supports the findings of this study is available 
from the corresponding author upon reasonable request.

% =========== Supplementary Materials =============
\clearpage
\linespread{1.2}
\setlength{\baselineskip}{15pt}

\newpage
\clearpage
\onecolumngrid
\mbox{}

\setcounter{section}{0}
\setcounter{figure}{0}
\setcounter{table}{0}
\setcounter{equation}{0}
\setcounter{table}{0}
\setcounter{page}{1}

\renewcommand{\thesection}{\normalsize{Supplementary Note \arabic{section}}}
\renewcommand{\theequation}{S\arabic{equation}}
\renewcommand{\thefigure}{S\arabic{figure}}
\renewcommand{\thetable}{S\arabic{table}}
\renewcommand{\thepage}{S\arabic{page}}

%======================
\begin{center}
{\large Supplementary Materials for}

$\,$\\
\textbf{\large{Universal Magnetocaloric Effect near Quantum Critical Point of\\ Magnon Bose-Einstein Condensation}}

$\,$\\
Xiang \textit{et al.}
\end{center}

\textbf{This PDF file includes:}\\
Supplementary Text\\
Figures S1 to S10\\
Table S1\\

\newpage
%=====================

\section{Sample preparation and characterization}
\begin{figure*}[htp]
\includegraphics[width=0.8\linewidth]{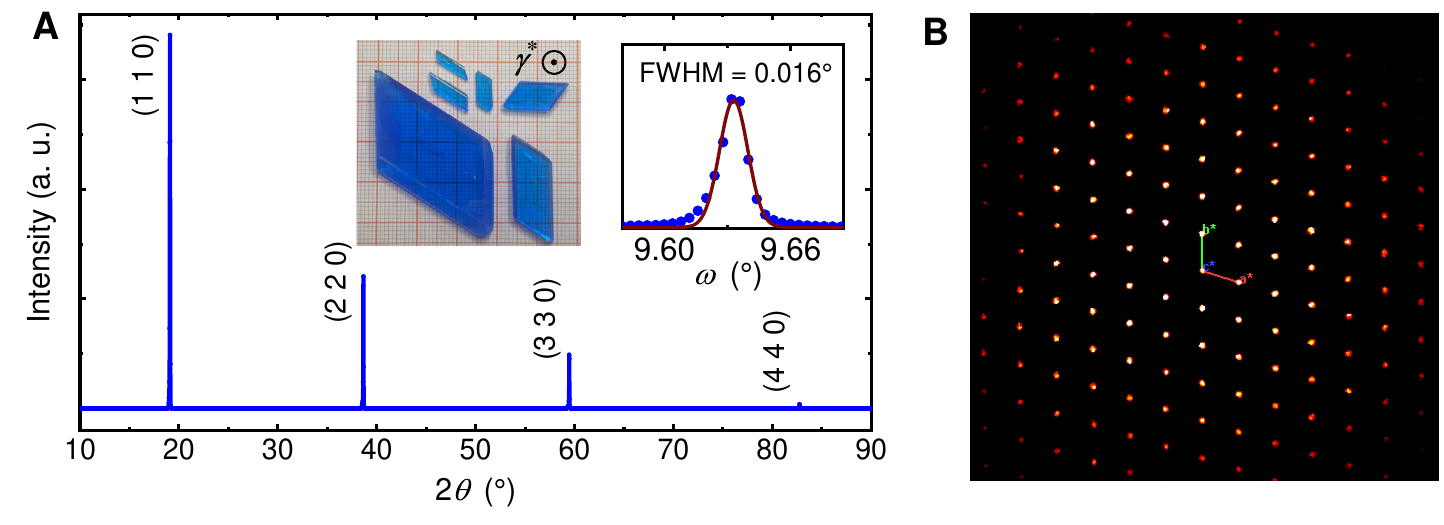}
\centering
\caption{\textbf{The XRD pattern and Laue diffractogram of \CSO single crystal.}
\textbf{(A)} {The $\theta$-2$\theta$ scan} recorded for the naturally grown surface {of the crystal} at room temperature 
by a Rigaku SmartLab diffractometer. The insets show the photo of CSO single crystals, and the rocking-curve 
scan of (110) reflection. The full width at half maximum (FWHM) of the peak is 0.016$^{\circ}$, indicating very high 
quality of the single crystal.
\textbf{(B)} The Laue diffractogram on the 001 {reciprocal-space} plane measured on a Rigaku XtaLAB Synergy-S diffractometer. {Sharp and} regularly spaced spots confirm good quality of the single crystal.}
\label{Fig_SM_XRD}
\end{figure*}

Single crystals of \CSO (CSO) were grown using commercial copper sulfate pentahydrate granules ($99.99\%$ purity). 
The seed crystals ($1 \times 1 \times 0.5$ mm$^3$ in size) had been prepared by cooling  supersaturated solution of 
$50$$^\circ$C to room temperature. Based on these seeding agents, the above steps were repeated in a vibration-free 
environment to obtain centimeter size single crystals. The CSO single crystal, belonging to the space group $P\overline{1}$, 
is a typical triclinic system. The lattice parameters are determined as $a$ = 6.1154~\AA, $b$ = 10.7126~\AA, $c$ = 5.9585~\AA, 
$\alpha$ = 82.399~$^\circ$, $\beta$ = 107.343~$^\circ$, $\gamma$ = 102.598~$^\circ$, with the Bruker D8 Venture x-ray 
diffractometer. The direction perpendicular to the naturally grown surface is identified as the $(110)$ crystallographic orientation 
by high-resolution X-ray diffraction (XRD) (fig.~\ref{Fig_SM_XRD}\textbf{A}) and defined as the $\gamma^*$-axis throughout 
this paper. The magnetic field is consistently applied parallel to this $\gamma^*$-axis in all measurements of this work. The 
high quality of the single crystals were confirmed by rocking-curve scan (see inset of fig.~\ref{Fig_SM_XRD}\textbf{A}) and 
Laue diffractogram, as shown in fig.~\ref{Fig_SM_XRD}\textbf{B}. 

\section{Density functional theory calculations}
The crystal structure of CSO hosts two nonequivalent magnetic sub-lattices — Cu-I and Cu-II ions — as 
established by prior thermodynamic and dynamic measurements~\cite{Miedema1962MagTherm,Groen1980NMR_Finite,
Mourigal2013}. However, systematic density functional theory (DFT) studies elucidating these dual magnetic 
systems remain scarce. Here, we present comprehensive DFT calculations on CSO, focusing on its magnetic 
structures and exchange couplings.

\begin{figure*}[htp]
\includegraphics[width=0.95\linewidth]{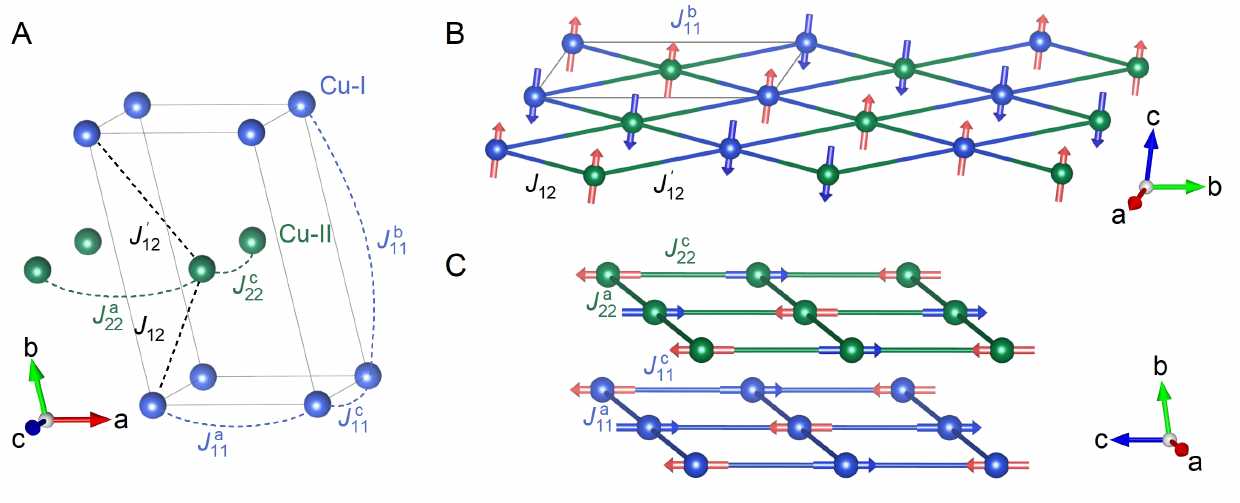}
\centering
\caption{\textbf{DFT calculations on \CSO.}
\textbf{(A)} The magnetic primitive cell of CSO. The blue (green) balls represent the 
Cu-I (Cu-II) ions, and the dashed lines indicate the spin couplings between Cu ions. 
\textbf{(B,C)} The structures of the $3 \times 3\times 1$ supercell and $3 \times 
1\times 3$ supercell. The arrows represent the magnetic moments of Cu ions.
% \textbf{(\textsc{D})} lists the calculated spin coupling values between Cu ions in CSO.
}
\label{Fig_SM_DFT}
\end{figure*}

We employ the Perdew-Burke-Ernzerhof exchange-correlation potential in the generalized gradient approximation 
(GGA) to perform the DFT calculations, with the experimental lattice parameters in Note~1.
 Figure~\ref{Fig_SM_DFT}\textbf{A} shows a magnetic primitive cell containing one Cu-I (blue ball) 
and one Cu-II ion (green ball). 
We denote the several (inequivalent) nearest-neighbor interactions as $J_{\alpha \beta}^{\gamma}$, where $\alpha, 
\beta = 1, 2$ correspond to the Cu-I, Cu-II ion respectively and $\gamma$ indicates the coupling directions. 
Based on a supercell, we compute total energies by considering four collinear spin alignments on two 
chosen sites. The corresponding coupling $J_{\alpha \beta}^{\gamma}$ is then determined using the standard 
four-state method~\cite{fourStatePRB} 
\begin{equation}
	J_{\alpha \beta}  = \frac{E_1 (\uparrow,\uparrow) + E_4 (\downarrow, \downarrow) 
	- E_2 (\uparrow, \downarrow) - E_3 (\downarrow, \uparrow) }{4 \times S_{\alpha} \times S_{\beta}},
\end{equation}
where $E_1$, $E_2$, $E_3$ and $E_4$ are the energies of the four spin configurations derived from DFT static 
calculations, and here $S_{\alpha} = S_{\beta} =1/2$ is used. 
Furthermore, we calculate the exchange couplings $J_{11}^{b}$, $J_{12}$ and $J_{12}'$ using a 
$3\times 3\times 1$ supercell (see fig.~\ref{Fig_SM_DFT}\textbf{B}), and determine the values of 
$J_{11}^a$, $J_{11}^c$, $J_{22}^a$ and $J_{22}^c$ through calculations performed on a 
$3\times 1\times 3$ supercell (see fig.~\ref{Fig_SM_DFT}\textbf{C})

From the DFT calculations, we identify two distinct magnetic sub-lattices in CSO. Cu-I ions form the weakly coupled 
spin chain structure along $a$-axis, while Cu-II ions are nearly free with much smaller magnetic couplings. From the 
results in Tab.~\ref{Tab:DFT}, we find that the dominate coupling $J_{11}^a\simeq 0.331$~meV along the $a$-axis, 
consistent with the spin-chain exchange strength of 0.252~meV determined from neutron scattering~\cite{Mourigal2013}. 
The other interactions, e.g., the couplings between Cu-II ions, are much smaller. 

\begin{table}[h!]
	\caption{Coupling strengths in \CSO calculated using the four-state method. }
    \centering
    \begin{tabular}{|p{4cm}|p{1.2cm}|p{1.2cm}|p{1.2cm}|p{1.2cm}|p{1.2cm}|p{1.2cm}|p{1.2cm}|} 
    \hline
    \centering{\textbf{$J_{\alpha \beta}^{\gamma}$}} & {$J_{11}^a$} & {$J_{11}^b$} & {$J_{11}^c$} & {$J_{22}^a$} & {$J_{22}^c$} & {$J_{12}$} & {$J_{12}'$} \\ 
    \hline
    \centering{\textbf{Coupling strength (meV)}} & {0.331} & {$\approx$ 0} & {0.007} & {-0.018} & {0.010} & {-0.025} & {$\approx$ 0} \\ 
    \hline
    \end{tabular}
    \label{Tab:DFT}
\end{table}

\section{Specific heat and magnetic entropy results}
\begin{figure*}[htp]
\includegraphics[width=0.9\linewidth]{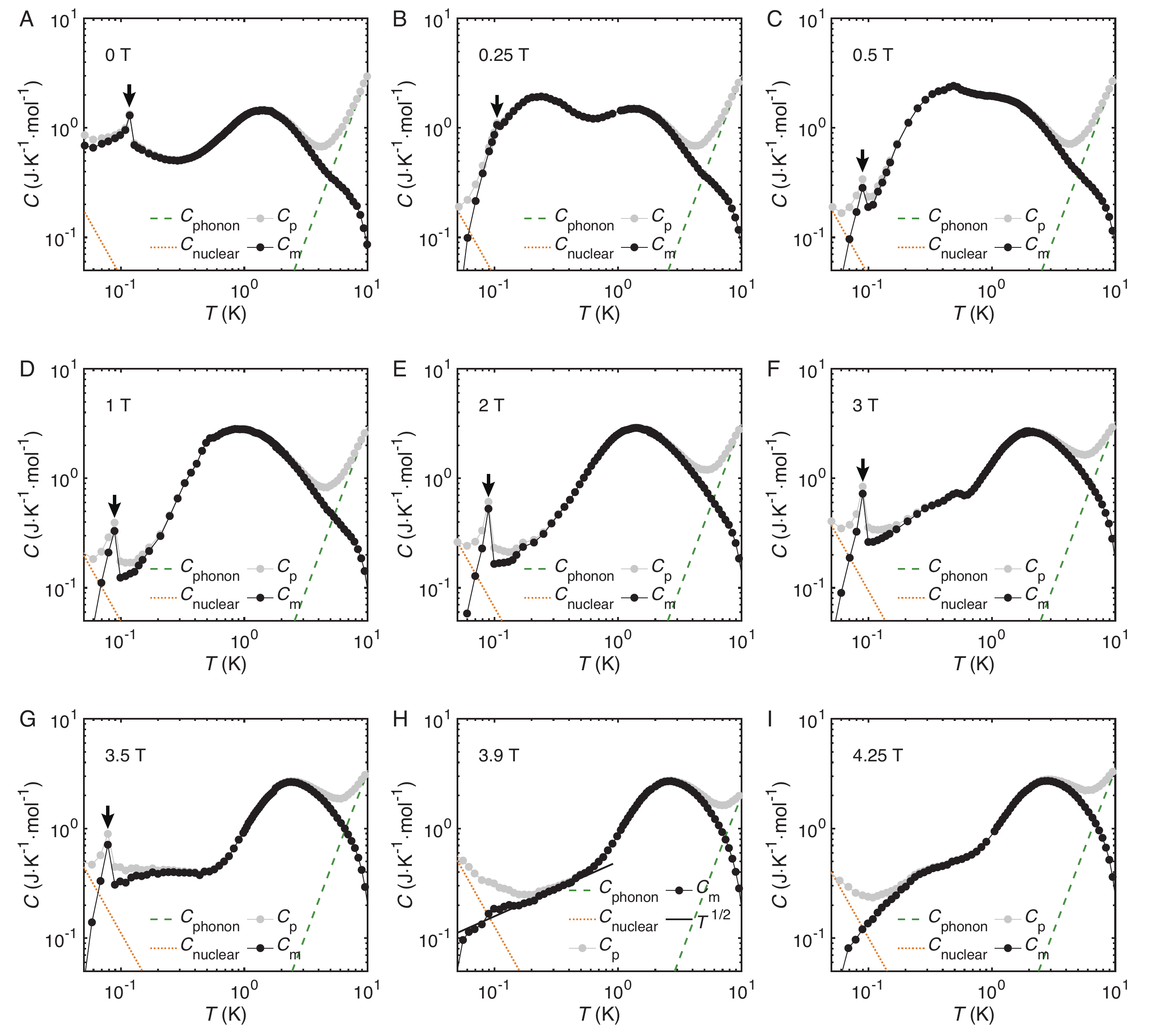}
\centering
\caption{\textbf{Specific heat data of \CSO for various magnetic field.}
The dark-colored dots represent the raw data $C_{\rm p}$, while the light-colored dots are the deduced magnetic specific heat 
$C_{\rm m}$ by subtracting the nuclear (dotted lines) and lattice phonon (dashed lines) contributions. 
Black arrows mark $\lambda$-transition temperatures. }
\label{Fig_SM_Cm}
\end{figure*}

The specific heat of CSO are measured down to 50~mK on a single crystal of 0.35~mg, 
using a commercial Quantum Design Physical Property Measurement System (PPMS)
 with a dilution refrigerator (DR) insert. The magnetic field is applied 
along the $\gamma^*$-axis (i.e., perpendicular to the naturally grown surface of the single crystal). 
At high temperatures (above about 10~K), the specific heat is mainly contributed by the lattice 
phonons, as described by the Debye model, 
\begin{equation}
	C_{\rm phonon} = 9N_A k_B \left(\frac{T}{\Theta_D}\right)^3 \int_{0}^{\Theta_D/T}
	\frac{x^4 e^x}{\left(e^x-1\right)^2} {\rm d}x,
	\label{C_phonon}
\end{equation}
where $N_A$ is the Avogadro constant and $k_B$ is the Boltzmann constant. The Debye temperature 
$\Theta_D$ is fitted by high-temperature data. At ultra-low temperatures ($T \lesssim $ 50~mK), 
the specific heat $C_{\rm p}$ exhibits a noticeable upward turn, which we attributed to the nuclear spin contribution. 
In our experimentally relevant temperature regime, the nuclear spin specific heat can be described by a 
high-$T$ expression, i.e., 
\begin{equation}
	C_{\rm nuclear} = \frac{a B_{\rm eff}^2}{T^2},
\end{equation}
where $a$ is a constant coefficient and $B_{\rm eff} = \sqrt{B^2+b^2}$ is the effective external field~\cite{Nuclear2014}. 
By fitting the low-temperature specific heat data at various magnetic fields, the coefficients are determined as $a = 0.055 $ 
mJ$\cdot$K$\cdot$mol$^{-1}\cdot$T$^{-2}$ and $b=2.8$~T. Consequently, the magnetic specific heat can be obtained using the relation 
\begin{equation}
        C_{\rm m} = C_{\rm p} - C_{\rm phonon} - C_{\rm \rm nuclear}. 
\end{equation}
Figure~\ref{Fig_SM_Cm} illustrates specific heat data of some typical fields. Dark-colored dots represent raw data $C_{\rm p}$, 
while light-colored dots are magnetic specific heat $C_{\rm m}$.

\section{Thermometer calibration in the adiabatic demagnetization measurements}
\begin{figure*}[htp]
\includegraphics[width=0.5\linewidth]{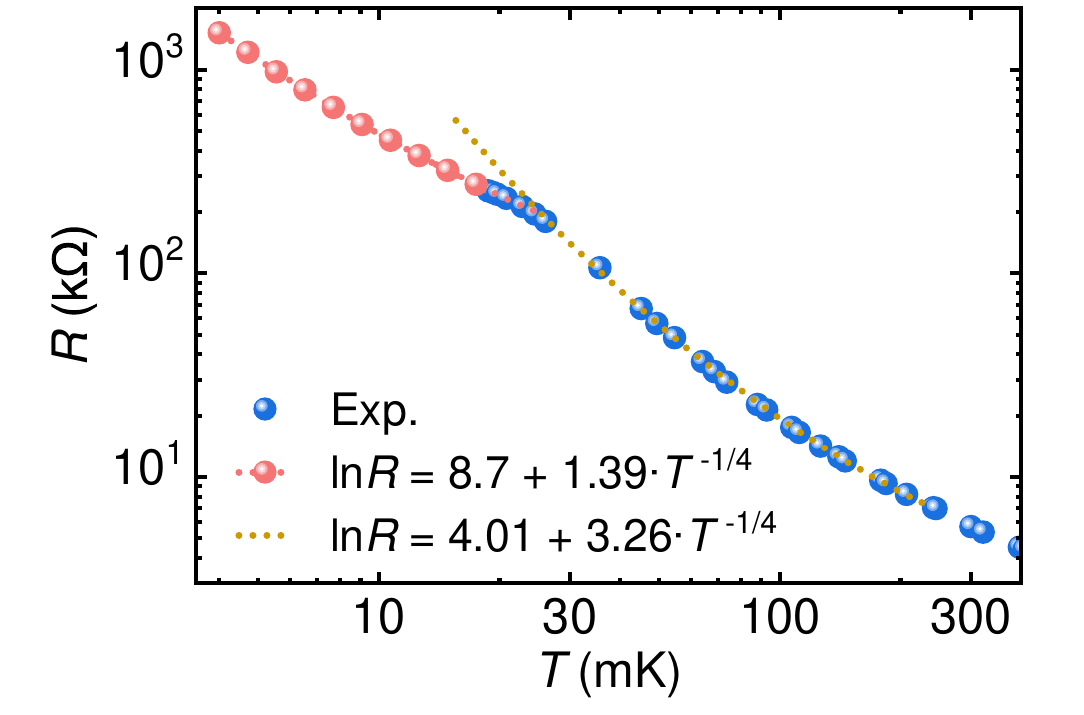}
\centering
\caption{\textbf{Calibration and extrapolation of thermometer.} 
The ruthenium oxide thermometer calibration is presented, with experimental measurements down to 18.75~mK 
shown as blue data points. High-temperature fits are displayed as a yellow dotted line, while 
the low-temperature calibration (extrapolated down to 4~mK) appears in red. 
}
\label{Fig_SM_TMCal}
\end{figure*}

The ruthenium oxide (RuO$_2$) thermometer is calibrated down to 18.75~mK at the zero field and 50~mK 
under magnetic field. Figure~\ref{Fig_SM_TMCal} illustrates the calibration data (blue dots) at the zero field. 
As temperature decreases, the resistance increases and reaches its maximum value 253~k$\Omega$ 
at 18.75~mK. At low temperatures, the resistance $R$ exhibits universal temperature dependence 
following $R = a \cdot \exp{(b \cdot T^{-1/4})}$
%, with $R_0=1170\,\Omega$ a temperature-independent constant
\cite{willekers1990thick}. For lower temperatures ($T<18.75$~mK), we extrapolate the 
calibration curve (indicated by the red solid dots in fig.~\ref{Fig_SM_TMCal}) to estimate the temperature.

\section{Thermal relaxation processes}

\begin{figure*}[htp]
\includegraphics[width=0.6\linewidth]{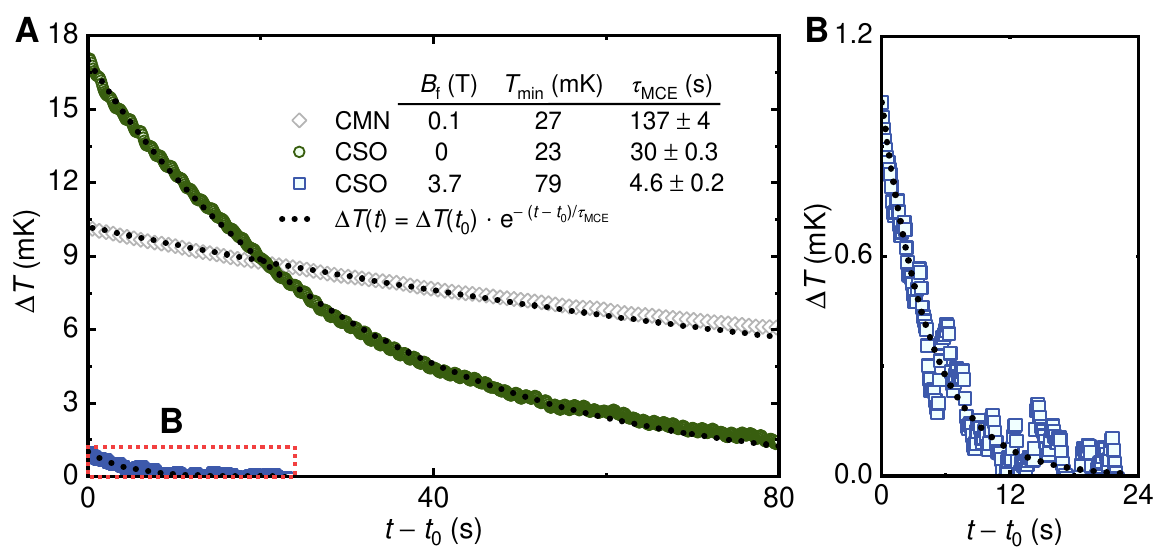}
\centering
\caption{\textbf{The magnetocaloric relaxation processes.} 
\textbf{A} The experimental data of thermal relaxations under different temperatures and magnetic fields. 
The fittings are $\Delta T(t) = \Delta T(t_0) e^{-(t-t_0)/\tau_{\rm MCE}}$, where $\Delta T(t) \equiv T(t) - T_{\rm min}$, 
$t_0$ is the initial time, $T_{\rm min}$ is the lowest (equilibrium) temperature, and $\tau_{\rm MCE}$ is the fitted 
thermal relaxation time. \textbf{B} shows a zoomed-in view of the red-dashed rectangular area denoted in \textbf{A}.}
\label{Fig_SM_RT}
\end{figure*}

\section{The nuclear magnetic resonance measurement}

\begin{figure*}[htp]
\includegraphics[width=0.45\linewidth]{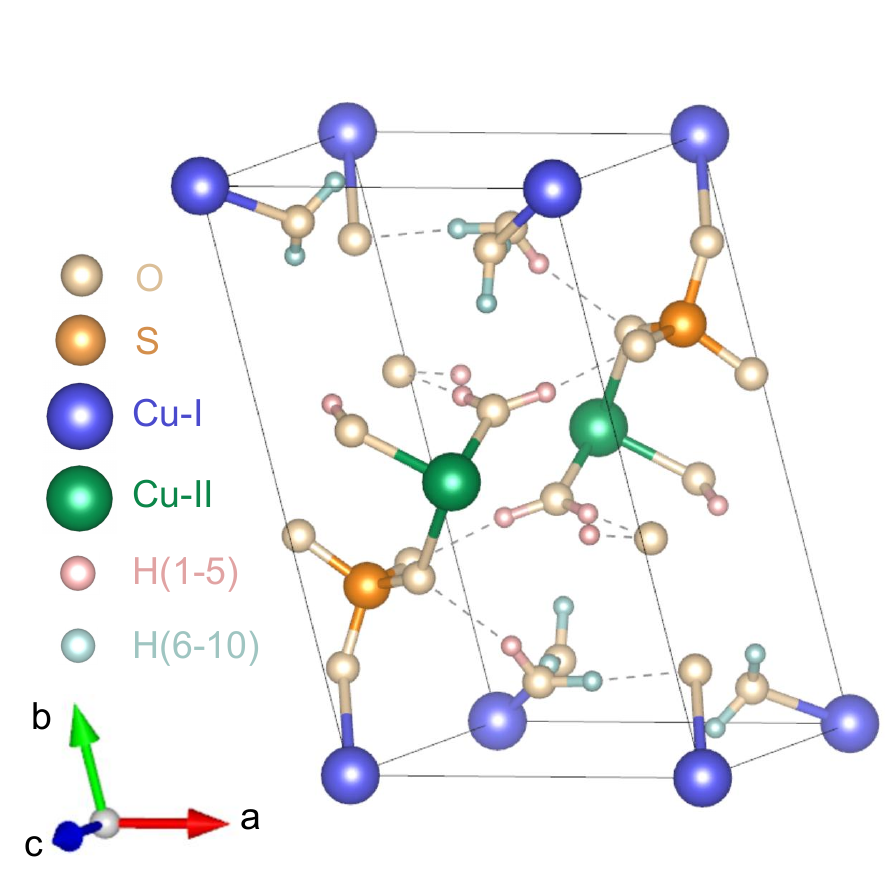}
\centering
\caption{\textbf{The crystal structure of \CSO.}
Blue balls represent the Cu-I ions, while green balls represent the paramagnetic Cu-II ions. There are a total 
of ten distinct hydrogen sites, namely H(1-5) close to Cu-II and H(6-10) sites close to Cu-I, respectively.
}
\label{Fig_SM_NMR_structure}
\end{figure*}

\begin{figure*}[htp]
	\includegraphics[width=0.6\linewidth]{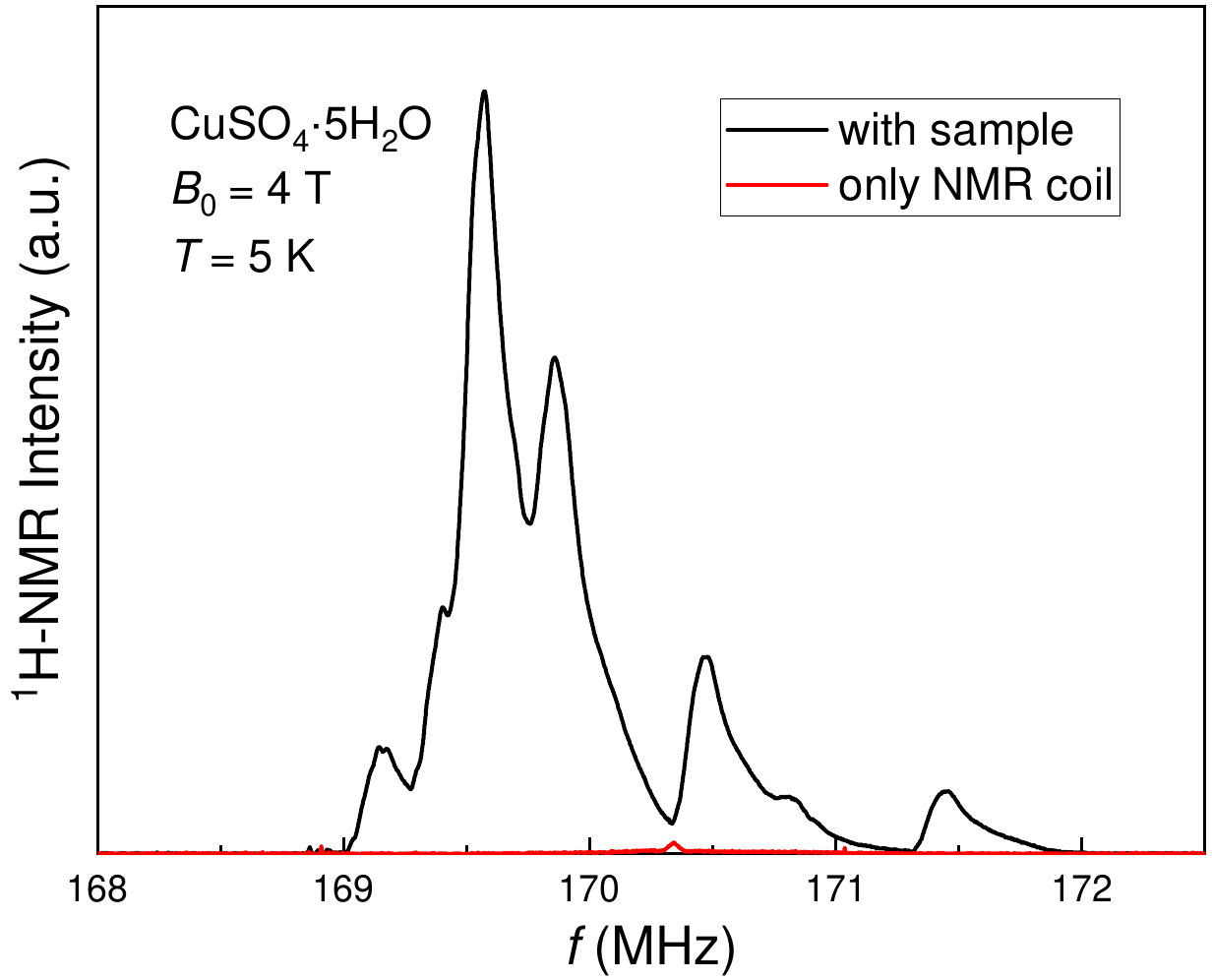}
	\centering
	\caption{\textbf{Comparison of $^1$H NMR spectra with and without \CSO sample in the coil.}
		The NMR spectra of the \CSO sample within the coil (the black line) and of only the coil (the red line) are measured under identical conditions ($B_0$ = 4 T and $T$ = 4 K). As nearly no NMR signal was observed when the sample was not in the coil, demonstrating that all the detected $^1$H NMR signals are indeed from the sample.
	}
	\label{Fig_SM_NMR_coil}
\end{figure*}

\begin{figure*}[htp]
	\includegraphics[width=0.6\linewidth]{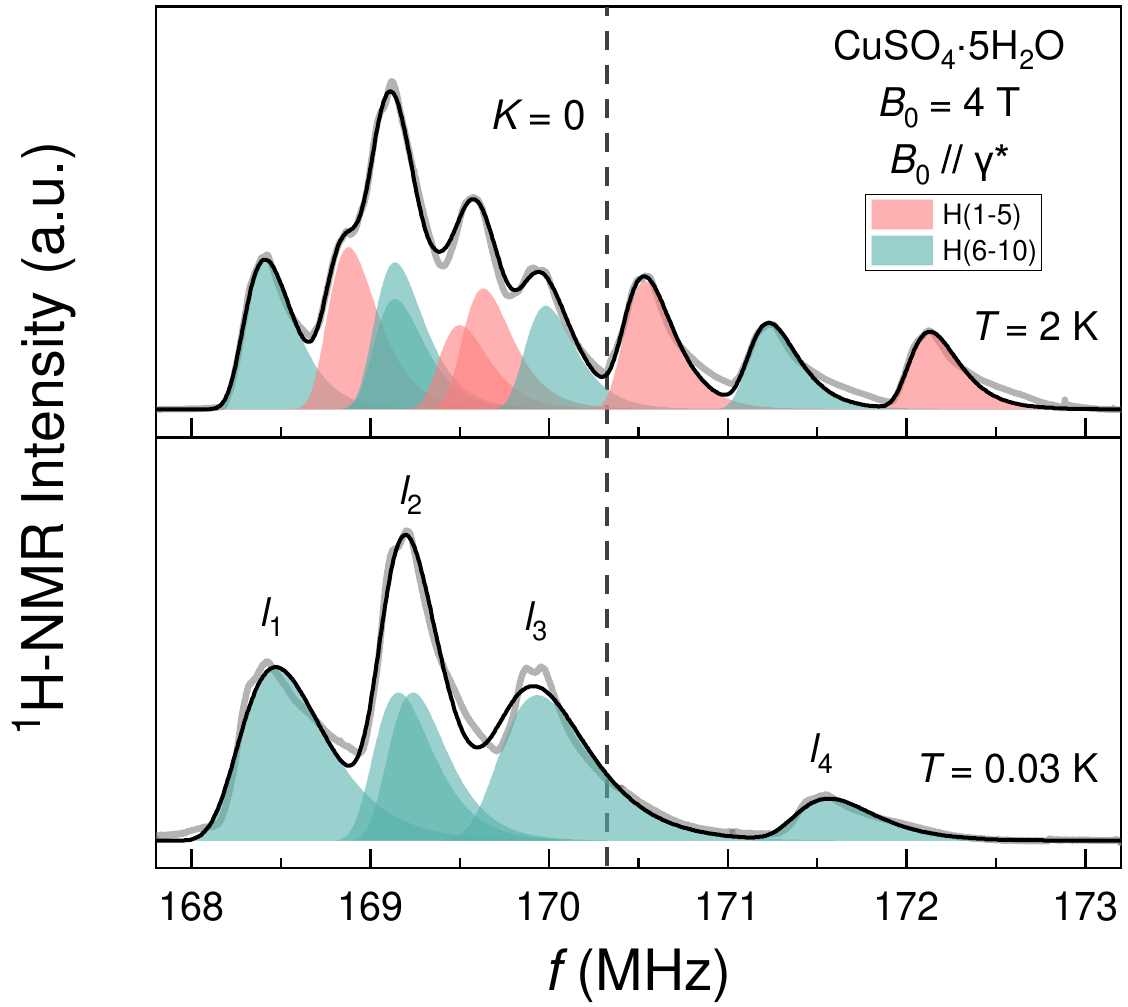}
	\centering
	\caption{\textbf{$^1$H NMR spectra of \CSO.}
		\textbf{} The upper and bottom panels present the spectra at 2 K and 0.03 K, respectively.  Owing to possible imperfect crystalline orientation of the sample, the NMR lines exhibit an slightly asymmetric line shape. Hence, we used extreme value distribution functions to fit the NMR spectra (solid lines)\cite{Asymline2017}. As shown in Fig.~\ref{Fig_SM_NMR_structure}, there are ten distinct hydrogen sites in \CSO. Since the nuclear spin of $^1$H is 1/2, ten lines should be expected to be observed in the NMR spectrum, as demonstrated by the previous NMR study~\cite{Poulis1951NMR}. Indeed, ten peaks with comparable spectral weight are observed at 2 K (see the upper panel), whereas only four lines ($l_1$ to $l_4$) corresponding to 5 peaks are observed at 0.03 K (see the bottom panel). Based on the previous NMR study~\cite{BOS1984269}, we can assign the NMR lines, as the red and green patterns correspond to the peaks from the H(1-5) and H(6-10) sites. The $^1$H nucleus of the H(1-5) sites has a stronger dipolar coupling with the paramagnetic copper ions Cu-II compared to that of the H(6-10) sites~\cite{WITTEKOEK1968293}. Thus, the spin-spin relaxation rate 1/$T_2$ of the $^1$H nucleus at the H(1-5) sites would be larger than that at the H(6-10) sites. 
		Meanwhile, the extremely small interaction of paramagnetic Cu-II ions (< 0.004 meV)~\cite{Mourigal2013}, which can induce an energy gap in the paramagnetic state under this magnetic field~\cite{Mukhopadhyay2012,Batista2014RMP}. Therefore, the 1/$T_1$ of the $^1$H nucleus at the H(1-5) sites would be extremely prolonged at very low temperatures. All these would  result in a reduction of the H(1-5) NMR signals at low temperatures.
		Therefore, the observed NMR lines are mainly contributed from the H(6-10) sites at low temperatures (see the bottom panel). We notice that the $l_1$, $l_2$ and $l_4$ lines exhibit a larger shift than the $l_3$ line, suggesting that the H sites corresponding to the $l_1$, $l_2$ and $l_4$ lines have a larger hyperfine coupling to Cu-I.
	}
	\label{Fig_SM_NMR_spectra}
\end{figure*}

\clearpage

\begin{figure*}[htp]
	\includegraphics[width=1\linewidth]{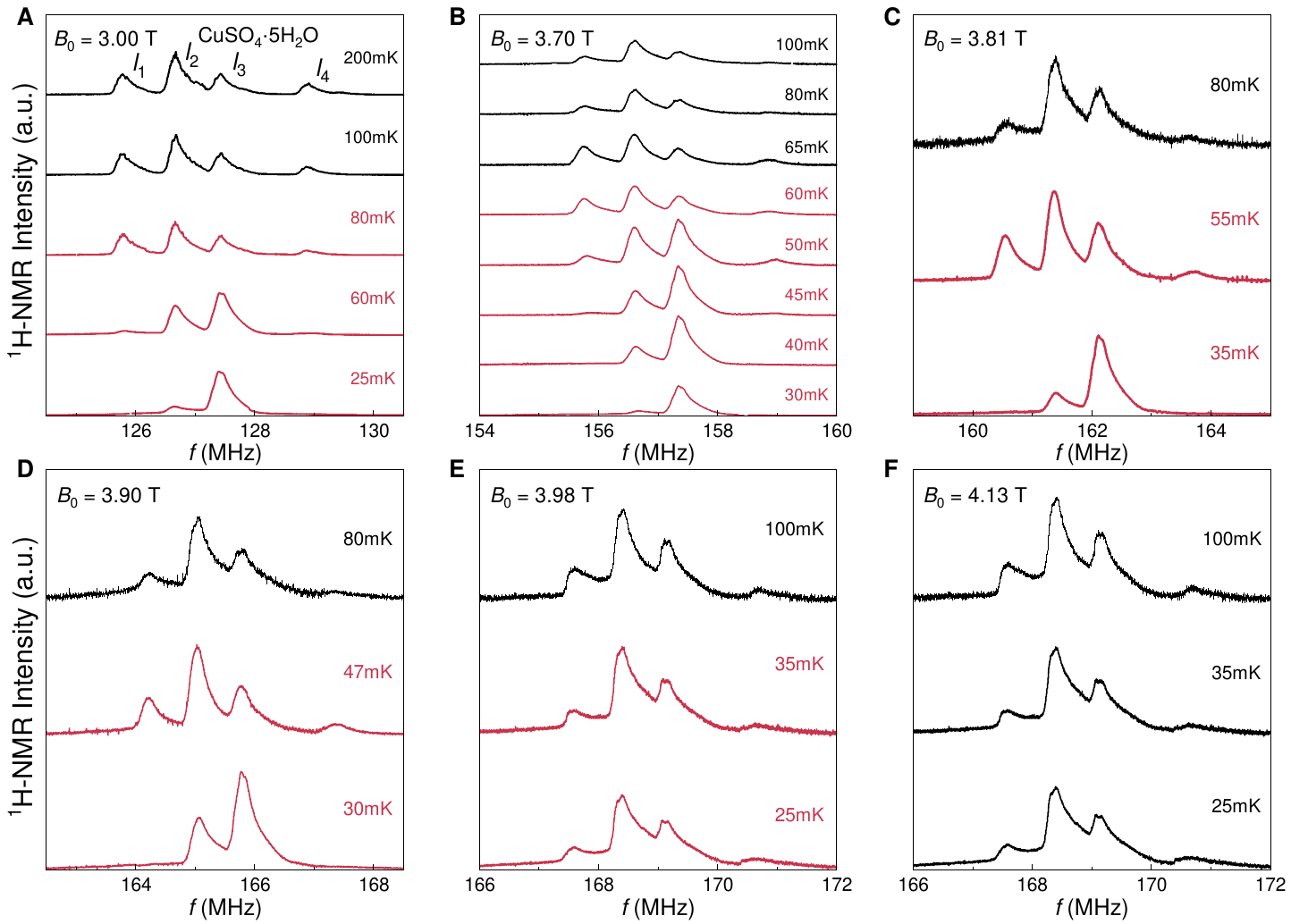}
	\centering
	\caption{\textbf{NMR spectral evidence for the phase transition.}
		\textbf{A-F}, the temperature dependence of $^1$H-NMR spectra at various fields. The black and red peaks are the NMR spectra in the normal state and ordered state, respectively. In the 3.00 T $\leq B \leq$ 3.90 T region, four lines are observed above $T_c$, while the intensities of $l_1$, $l_2$ and $l_4$ decrease rapidly below $T_c$. The reduction of the intensity implies a shift in spectral weight, which is a typical characteristic of magnetic phase transitions. For $B \geq$ 3.98 T, the change in spectra is scarcely observable even below $T_c$. Hence, we measured the spin-lattice relaxation rate 1/$T_1$ at both $l_2$ and $l_3$ lines to further determine the transition temperature $T_c$.
	}
	\label{Fig_SM_NMR_all}
\end{figure*}

\begin{figure*}[ht]
\includegraphics[width=0.55\linewidth]{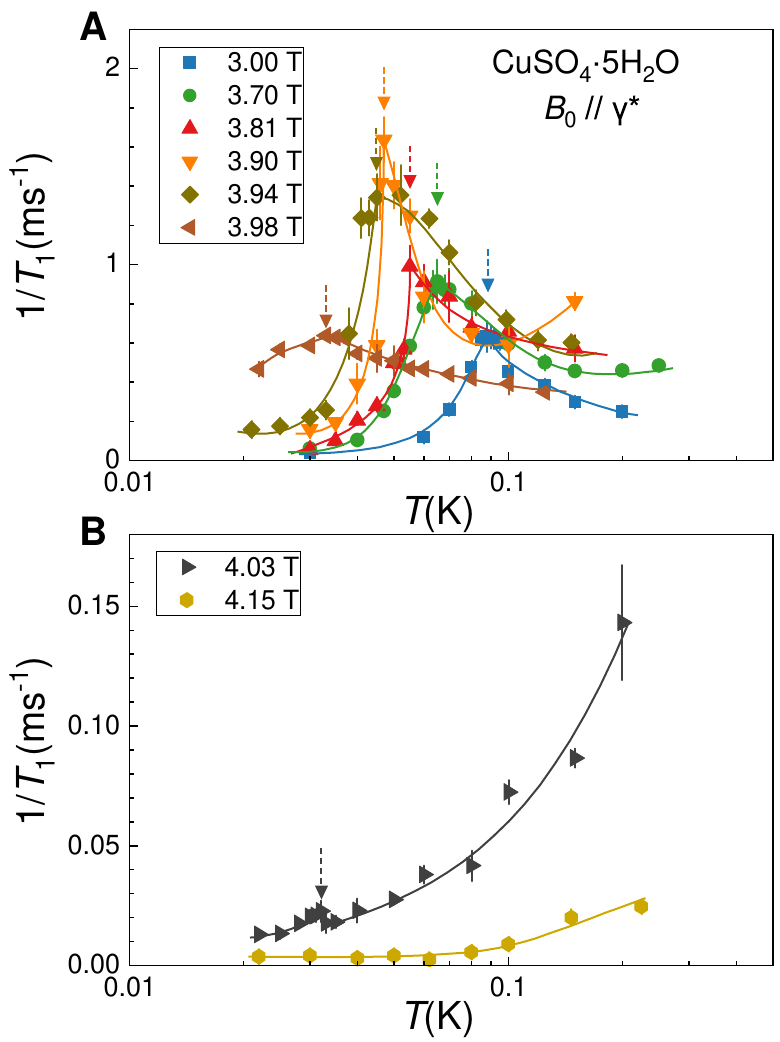}
\centering
\caption{\textbf{1/$T_1$ evidence for the phase transition.}
\textbf{A} and \textbf{B} are the temperature dependence of 1/$T_1$ measured at the $l_3$ line for various fields. Similar to the temperature dependence of 1/$T_1$ measured at the $l_2$ line (see Fig.~\ref{Fig2} in the main text), 1/$T_1$ drops rapidly below the transition temperature $T_c$. Generally, $1/T_{1}T$ is proportional to the imaginary part of the dynamic susceptibility perpendicular to the applied field ($\chi''_\perp(q,\omega)$) and can be expressed as $\frac{1}{T_1T} \propto \sum_q |A(\textbf{q})|^2 \frac{\chi''_\perp(\textbf{q},\omega)}{\omega}$~\cite{Moriya1963Effect}. $A(\textbf{q})$ is the wave vector \textbf{q} dependent hyperfine coupling constant and $\omega$ is the NMR frequency. The Knight shift $K$ probes the static magnetic susceptibility at \textbf{q} = 0, while $1/T_{1}$ is also sensitive to the spin fluctuations at $\textbf{q} \neq 0$.  For the antiferromagnetic spin fluctuations, there will be a peak around a finite wave vector. Hence, $A(\textbf{q})$ in $1/T_{1}T$ is significant for all H sites corresponding to $l_1$-$l_4$ lines, although  the Knight shift of $l_3$ line is much smaller than that of other lines.
	}
	\label{Fig_SM_P2_T1}
\end{figure*}

\clearpage

\end{document}